\documentclass{mn2e}
\usepackage{graphicx}
\usepackage{subfigure}
\newcommand{\e}{\begin{equation}}
\newcommand{\dd}{\mathrm d}
\newcommand{\st}{\end{equation}}
\newcommand{\m}{\begin{displaymath}}
\newcommand{\n}{\end{displaymath}}
\newcommand{\msun}{M_\odot}

\begin{document}
\title[Hypervelocity Stars from the Andromeda Galaxy]{Hypervelocity Stars from the Andromeda Galaxy}

\author[B. D. Sherwin, A. Loeb and R. M. O'Leary]{Blake D. Sherwin$^{1,2}$\thanks{bds30@cam.ac.uk}, Abraham Loeb$^1$\thanks{aloeb@cfa.harvard.edu} and Ryan M. O'Leary$^1$\\
$^{1}$ Harvard-Smithsonian Center for Astrophysics, 60 Garden Street, Cambridge, MA 02138, USA\\
$^{2}$ Department of Applied Mathematics and Theoretical Physics, University of Cambridge, Cambridge CB3 0WA, UK}

\maketitle

\begin{abstract}
Hypervelocity stars (HVSs) discovered in the Milky Way
(MW) halo are thought to be ejected from near the massive black hole
(MBH) at the galactic centre. In this paper we investigate the spatial and
velocity distributions of the HVSs which are expected to be similarly
produced in the Andromeda galaxy (M31). We consider three different HVS
production mechanisms: (i) the disruption of stellar binaries by the
galactocentric MBH; (ii) the ejection of stars by an in-spiraling
intermediate mass black hole; and (iii) the scattering of stars off a
cluster of stellar-mass black holes orbiting around the MBH. While the first two
mechanisms would produce large numbers of HVSs in M31, we show that the
third mechanism would not be effective in M31. We numerically calculate
$1.2 \times 10^6$ trajectories of HVSs from M31 within a simple model of
the Local Group and hence infer the current distribution of these stars. Gravitational focusing of the HVSs by the MW and the diffuse Local Group medium leads to high densities of low mass ($\approx 1 \msun$) M31 HVSs near the MW. Within the virialized MW halo, we expect there to be of order $1000$ HVSs for the first mechanism and a few hundred HVSs for the second mechanism; many of these stars should
have distinctively large approach velocities ($<- 500$$\ \mathrm{km} \ \mathrm{s}^{-1}$). In addition, we predict $\sim 5$ hypervelocity RGB stars within the M31 halo which could be identified observationally. Future MW astrometric surveys or searches for distant giants could thus find HVSs from M31.
\end{abstract}

\section{Introduction} 

Hypervelocity stars (HVSs) travel at such high speeds that they are not
gravitationally bound to the galaxy from which they originate. The first HVS was discovered in the Milky Way (MW) halo by Brown et al.\ (2005); nine
other HVSs in the MW halo have since been found by Edelmann et al.\
(2005), Hirsch et al.\ (2005) and Brown et al.\ (2006a,b, 2007b). By selection,
most of the observed stars are B-type stars. The HVSs have typical speeds of $550$$\ \mathrm{km} \ \mathrm{s}^{-1}$. There is also evidence for a bound population of stars similar to the
HVSs but with speeds slightly lower than the escape velocity of the MW
(Brown et al.\ 2007a). The observations of MW HVSs indicate that all but one of these stars
originate from the centre of the galaxy\footnote{The HVS discovered by Edelmann et al.\ (2005) appears to originate from the Large Magellanic Cloud.}. This is easily understood if HVSs
are ejected by the massive black holes (MBHs) which inhabit the centres of
galaxy bulges.

The ejection of HVSs from galactocentric MBHs was first predicted by Hills
(1988). The proposed mechanism was the disruption of a tight stellar binary
by the MBH, which results in the capture of one of the stars and the
ejection of the other as a HVS. This tidal breakup (``TB'') mechanism has
been further investigated by Yu \& Tremaine (2003), Gualandris et al.\
(2005), Ginsburg \& Loeb (2006), and Bromley et al.\ (2006). Another model for the production of HVSs
is the scattering of stars bound to the galactic MBH by an intermediate or high mass in-spiraling
black hole (``IBH'' mechanism - Yu \& Tremaine 2003;
Levin 2006; Baumgardt et al.\ 2006; Sesana et al.\ 2006, 2007b). Though the
results of Sesana et al.\ (2007a) indicate that this second model may be
inconsistent with the statistics of the observed MW HVSs, it cannot be
ruled out in other galaxies. The third mechanism for the ejection of HVSs was proposed
by O'Leary \& Loeb (2006); this involves the scattering of stars by a
cluster of orbiting stellar-mass black holes which have segregated around
the galactocentric MBH (``BHC'' mechanism). It has not yet been
conclusively determined which of these mechanisms is dominant in the MW,
and so we will consider them all in our work.

The HVSs originating from the MW centre have been studied in some detail by
both observers and theorists. However, the production of HVSs is of course
not unique to our galaxy; indeed it is expected to be a feature of all
galaxies with central MBHs. The HVSs which escape from other galaxies will
become, to quote Hills (1988), ``intergalactic tramps''. In particular, the
neighbouring Andromeda galaxy (M31) is fairly similar to the MW and has a
MBH at its centre (Bender et al.\ 2005). We thus expect a HVS population
similar to that observed in the MW to be produced in M31, with some of the
escaping stars gravitationally focused into the MW. Observations of such
stars could be used to determine the HVS production mechanism, elucidate
the properties and merger history of the core of M31, and constrain the
distribution of mass within the Local Group. This leads to two questions
which are the subject of our paper: {\it (i) what is the distribution of
stars ejected from M31?; and (ii) can we observe them from our galaxy?}

To answer these questions one needs to consider the history of the MW and
M31 within the Local Group of galaxies (of which they are the largest
constituents). Kahn and Woltjer (1959) first introduced the ``timing
argument'', by which the MW and M31 are assumed to have formed close to
each other in the early universe and were subsequently pulled apart by the
general cosmological expansion. The two galaxies are now decoupled from the
expansion and have traced out much of a Keplerian orbit, which explains
the current observed approach velocity of $\approx 120$$\ \mathrm{km} \ \mathrm{s}^{-1}$ (Binney \&
Tremaine 1987). Given the current approach velocity, the separation of the
galaxies, the age of the universe, and assuming that the Local Group has no
angular momentum, one can solve Kepler's equations to find an estimate of
the Local Group mass -- $\approx (3$--$5) \times 10^{12} \msun$ (see Binney
\& Tremaine 1987, Fich \& Tremaine 1991). The timing argument has since
also been applied to systems with angular momentum and realistic mass
distributions (see e.g.\ Peebles et al.\ 1989; Valtonen et al.\ 1993;
Peebles 1994; Peebles et al.\ 2001; Sawa \& Fujimoto 2005; Loeb et al.\
2005).

The timing argument also allows an approximation to be made about the
separation of the MW and M31 for the dynamics of HVSs. A feature of
the galaxies' eccentric orbit is that the galaxies spend a large
fraction of the orbital time near apogalacticon, with a relative speed less than the current value. Since the HVSs move at speeds which are at least a
few times larger than the current relative radial speed of M31 and the
MW, a reasonable approximation in following the HVS dynamics is to
assume a static configuration over the last $10^{10}$ years, with M31 and the MW separated by their current distance. However, we also verified that the HVS
distributions are not significantly affected by the addition of a
radial component to the HVS velocities of order $100$$\ \mathrm{km} \ \mathrm{s}^{-1}$.

Recent studies of the Local Group investigated the galactocentric transverse velocity of
M31. Peebles et al.\ (2001) applied the action principle to the motion of
galaxies within $20$ Mpc of the Local Group and obtained a transverse
velocity of $\approx 200$$\ \mathrm{km} \ \mathrm{s}^{-1}$. Loeb et al.\ (2005) analysed the proper
motion of M31's satellite galaxy M33 using numerical simulations and
estimated the transverse velocity of M31 as $\approx 100$$\ \mathrm{km} \ \mathrm{s}^{-1}$. A recent analysis of the motion of M31's satellite galaxies by van der Marel \& Guhathakurta (2007) gave an M31 transverse velocity of $\approx 40$$\ \mathrm{km} \ \mathrm{s}^{-1}$.In our work
we neglect the transverse velocity as it has only a small influence on the
trajectories of the fast HVSs.

In the following sections we examine the trajectories of HVSs from M31 within the
Local Group and estimate their current distribution and observability. We
first construct a simple model of the Local Group and its gravitational
field as described in \S$2.1$. We then numerically integrate the equations
of motion for a large number of possible HVS trajectories ($\approx
10^{6}$) to find the current HVS distribution within the Local Group,
sampling from assumed probability distributions of HVS ejection direction,
ejection speed, and stellar mass, as detailed in \S$2.2$. An analysis of
how to perform the simulations most efficiently and a discussion of the
numerical methods we use to find the HVS trajectories and distributions is
given in \S3. The results of our simulations are presented in \S4 and
discussed in \S5. We comment on the observational outlook for our results
in \S6 and finally conclude in \S7.

\section{Modelling the System}
\subsection{The Local Group}
The trajectories and the spatial distributions of the HVSs produced in M31
depend on the gravitational field within the Local Group. A number of
models have been constructed of the mass distribution within the entire Local Group (Cox \&
Loeb 2007) and of the MW and M31 (Klypin et al.\ 2002; Widrow \& Dubinski 2005; Seigar et al.\ 2006). Much of the Local Group's mass is contained within the MW and M31 halos. These galaxies are surrounded by a diffuse Local Group medium of
dark matter and gas. Based on cosmological simulations in which similar
systems form, the mass in the diffuse medium is expected be a substantial
fraction of the total mass (Cox \& Loeb 2007); indeed the timing argument's
mass estimate of $\approx (3$--$5) \times 10^{12} \msun$ is generically
larger than the sum of the MW \& M31 halo masses ($\sim 2.6 \times
10^{12}\msun$).

We will thus assume in our model that the mass in the intragroup medium is
equal within a factor $f$ to the mass contained within the virialized halos
of M31 and the MW, $M_A + M_M$. For simplicity, this diffuse Local Group
mass is taken to be uniformly distributed within a sphere of radius $R_G$
about the midpoint between M31 and the MW. The two galaxies are separated
by 780 kpc (McConnachie et al. 2005, Ribas et al. 2005). 

The mass of both the bulge and disc of M31 and the Milky Way is much
smaller than the masses of their dark matter halos, so the
gravitational influence of the bulge and disc is only important very
close to the galaxies' centres. As we are interested in approximate
HVS trajectories and distributions within the entire Local Group and
do not require precise trajectories near the galactic centre, we
neglect the gravitational field of the disc and bulge in our Local
Group model. However, the M31 disc and bulge significantly slow an
HVS which is escaping the galactic centre; we take this into account
by reducing the HVS initial velocity by an amount which corresponds to
the kinetic energy required to escape the disc and bulge. To estimate this
reduction in velocity, we assume a Hernquist bulge (Hernquist 1990)
with a mass $m_b=2 \times 10^{10} \msun$ (Klypin et al.\
2002) and a scale radius $a_b=1.8$ kpc (Widrow \& Dubinski
2005), as well as an exponential disc with a mass $m_d=8
\times 10^{10} \msun$ and a scale radius $R_d=5.5$ kpc
(Widrow \& Dubinski 2005). The central potential of the bulge is $-G
m_b/a_b$ (Widrow \& Dubinski 2005), and we calculate the central
exponential disc potential to be $-G m_d/R_d$. The escape velocity
$v_{bd}$ from the bulge and disk is thus $v_{bd}\approx 470\
\mathrm{km} \ \mathrm{s}^{-1}$; a HVS with an initial speed $v$ (once
outside the influence of the MBH) has a speed
$v_0=\sqrt{v^2-v_{bd}^2}$ after escaping the bulge and disk. We assume
that the gravitational field of the M31 disc does not modify the
initial angular distribution of HVSs as the disc's field is small and
mainly radial.

We model the dark matter distribution within
the MW \& M31 galaxy halos with the Navarro, Frenk and White (1996;
hereafter NFW) density profile which follows from cosmological galaxy
formation simulations: \e \rho(r)=\frac{\rho_0}{\left(c
r/R_{200}\right)\left(1+cr/R_{200}\right)^2}, \st where $c$ is the
concentration parameter, $R_{200}$ is the virial radius of the galaxy at
which we cut off the halos (interior to which the average density is 200
times the mean density of the Universe), and where $\rho_0$ is a
constant chosen to give the correct total virial mass $M_{A}$ (for M31) or
$M_{M}$ (for MW). The total density in our model is simply the
superposition of the densities of the two galaxies and the intragroup
medium.

\begin{figure}
	\centering
		\includegraphics[width=.49\textwidth,clip]{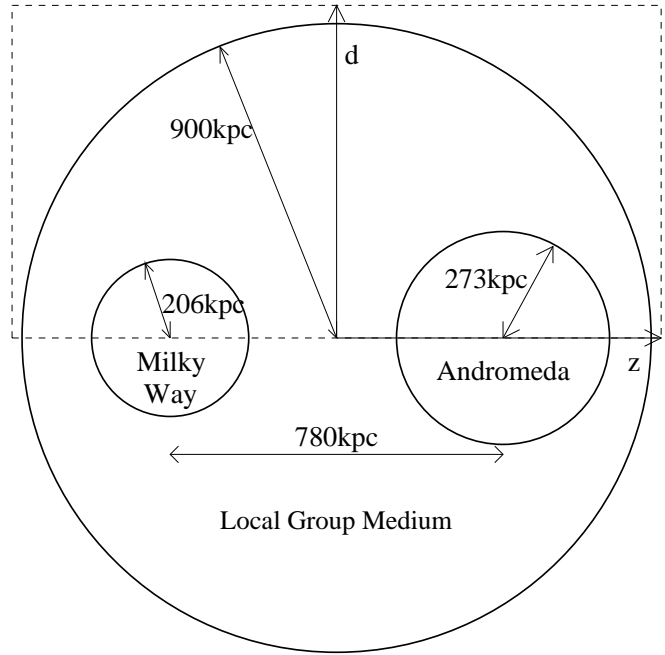}
		\caption{Our model of the Local Group. The dashed line
		indicates the range of our later plots.}
		\label{localgroup}
\end{figure}

Our Local Group model is thus completely specified by the parameters
$M_{A}$, $M_{M}$, $f$, $R_{A,200}$, $R_{M,200}$, $R_G$, $c_A$ and
$c_M$. The standard values of these constants which we use for our model
are $M_{A}=1.6\times 10^{12}\msun$, $M_{M}=1 \times 10^{12}\msun$, $f=1$, $c_A=c_M=12$, $R_{A,200}=273$ kpc, $R_{M,200}=206$ kpc and $R_G=0.9$
Mpc. These numbers are similar to those in Cox \& Loeb (2007), which are based on
observations and are also consistent with simulations of the development of
the Local Group. The model is depicted in Figure \ref{localgroup}.  In
following the dynamics of fast HVSs near the turnaround epoch of the Local
Group, we will assume that the configuration shown in Figure
\ref{localgroup} has not changed significantly over the age of the Local Group ($\sim 10^{10}$yr).

The simplicity of our model enables us to find analytic expressions for the
gravitational field, which increases the efficiency of our numerical
calculations. Making use of the spherical symmetry of all the features in
our model, the gravitational field can easily be calculated from Gauss'
law. Within the Local Group the
diffuse mass leads to the field \m \mathbf g_G( \mathbf
r)=\frac{-G(M_{A}+M_{M})f}{R_G^3}\ \mathbf r . \n The field within the MW
resulting from the NFW halo profile is a central field of magnitude (Navarro et al.\ 1996) \e
\label{NFW}
|\mathbf g_M(\mathbf r)|=\frac{G M_{M}}{R_{M,200}^2}\frac{1}{x^2}\frac{\ln(1+cx)-cx/(1+cx)}{\ln(1+c)-c/(1+c)} \st
where $\mathbf r_M$ is the position of the MW centre and $x=|\mathbf r -\mathbf r_M|/R_{M,200}$. The expression for the M31 field is analogous. The fields have a $1/r^2$ form outside the galaxy or
Local Group.

While the halo field is not singular at the centre of the galaxy, we avoid numerical difficulties caused by the denominator by including a softening length $\epsilon R_{M,200}$:
\begin{eqnarray}
\mathbf g_M(\mathbf r)
& = & -\frac{G M_{M}}{R_{M,200}^2}\frac{1}{(x+\epsilon)^2} \frac{\ln(1+cx)-cx/(1+cx)}{\ln(1+c)-c/(1+c)} \nonumber \\ & & \times \frac{\mathbf r -\mathbf r_M}{|\mathbf r -\mathbf r_M|+\epsilon R_{M,200}}.
\end{eqnarray}
We choose a small value of $\epsilon=0.01$ which introduces a realistic core in the region of the galactic bulge and is found not to influence trajectories away from the galaxy centres.

The total gravitational field is found by superposition: $\mathbf
g_t=\mathbf g_G+ \mathbf g_M+ \mathbf g_A$.

\subsection{The HVS Production Rate}

To accurately model the ejection of HVSs from M31, we must determine the
number of HVSs produced, the probability distribution of the HVSs' initial
velocities, and the distribution of their stellar masses. For simplicity,
we will assume that HVSs have been ejected in the same manner throughout
the lifetime of the Local Group; the velocity and mass probability
distributions are thus assumed constant over the last $10^{10}$ years, with
no HVS production before this time. We neglect the possibility that
multiple mechanisms simultaneously contribute to the HVS production and
only consider the mechanisms individually.

We first discuss the number of HVSs produced in the MW. A HVS
production rate in the MW can be estimated by extrapolating the observations of
hypervelocity blue stars in the MW halo to lower masses. We assume that star formation in the MW galactic centre
has followed a flat initial mass function $\frac{\dd N}{\dd m}$ as found in
current observations of the galactic centre, $\frac{\dd N}{\dd
  m}\propto m^{-0.85}$ for $m \geq 1\msun$ (Maness et al.\ 2007). The
cumulative distribution of stellar masses in the galactic centre
$P(m)$ is given by $P(m)\propto \frac{\dd N}{\dd m}t_l$ (see Alexander
\& Sternberg 1999, O'Leary \& Loeb 2007, Sesana et al.\ 2007a, Perets
et al.\ 2007). This implies that $P(m)\propto m^{-3.35}$, where $t_l$ is the
stellar lifetime, which is less than a Hubble time, and we have used
$t_l\propto m^{-2.5}$ (Kippenhahn \& Weigert 1990). Taking the masses
of ejected HVSs to be distributed according to $P(m)$ and assuming
that HVSs were produced at a constant rate, the HVS sample of
Brown et al.\ (2006a, 2007b) leads to a rough estimate of slightly more than
$10^6$ HVSs ejected from the MW centre over the age of the local
group.

The number of HVSs produced in M31 will not necessarily be the same as the
number produced in the MW. While M31 is overall very similar to the MW, the
masses of the MBHs in the galaxies are very different; the mass of the MW's
MBH (Sgr A$^*$) is $\approx 3.6 \times 10^6 \msun$ (Eisenhauer et al.\
2005), whereas the mass of the MBH in M31 is $\approx 1.4 \times 10^8
\msun$ (Bender et al.\ 2005). In addition, the stellar populations in the
centres of the two galaxies have different velocity dispersions, mass
functions and densities (Bender et al.\ 2005, Genzel et al.\ 2003). As the
MBH and its stellar environment play a large role in all HVS production
mechanisms, we must investigate how the differences will influence the
relative HVS production rates in the MW and in M31. We can then simply scale the rates in the MW (inferred from observations) for each of the mechanisms to obtain the total number of HVSs
produced in M31.

We briefly review some features of the stellar environment of MBHs. The
stars in the galactic centre have a typical mass $m_s\sim 1\msun$. The radius of
influence, $r_i$, is the distance to which the MBH significantly influences
the velocity and spatial distribution of stars; at this radius the kinetic
energy in the random motions of the stars (characterised by
the velocity dispersion $\sigma$) is roughly equal to the stars' potential
energy in the gravitational field of the MBH, so that $r_i
\approx GM_\bullet/\sigma_i^2$, where $M_\bullet$ is the mass of the central MBH. Outside the radius of influence, $\sigma$ is
approximately constant with radius ($\sigma\approx \sigma_i$); inside the
radius of influence, the velocity dispersion roughly equals the circular
velocity ($\sigma(r)\approx \sqrt{GM_\bullet/r}$). The density of
stellar mass near the black hole is assumed to be described inside the radius of influence by the
theoretically derived expression (Young 1980): 
\e \rho(r)=\rho_i
(r/r_i)^{-1.5}.
\label{rhoeq}
\st 
Outside the radius of influence, we assume an isothermal stellar distribution, so that
\e
\rho(r)=\rho_i(r/r_i)^{-2}.
\st
While one could also determine a density profile based solely on current observations and not on theoretical considerations, the current density would not necessarily correctly describe the density over the past $10^{10}$ years; we believe that the above theoretical expressions for the density should provide a better average model over the age of the Local Group. Our density profile agrees fairly well with observations of the galactic centres of both the
MW (Genzel et al. 2003) and M31 (Lauer et al. 1993). 
We use the following values of $r_i$, $\rho_i$ and $\sigma_i$ (assuming a
density profile as in Eq. \ref{rhoeq}): for the MW $r_M=1.66$pc, $\rho_M=6
\times 10^{4}\msun/$pc$^3$ and $\sigma_M\approx100$$\ \mathrm{km} \ \mathrm{s}^{-1}$ (Genzel et al.\
2003); for M31 $r_A=30$pc, $\rho_A=6\times 10^2\msun/$pc$^3$ and
$\sigma_A\approx150$$\ \mathrm{km} \ \mathrm{s}^{-1}$ (Bender et al.\ 2005, Lauer et al.\ 1993). We
assume a binary fraction $\eta$ of $\eta=0.1$.

In the BHC model, a cluster of stellar-mass black holes has segregated
close to the MBH. The stars near the MBH are moving at high speeds, so that
scattering of these stars off the orbiting stellar-mass black holes leads
to hypervelocity ejections. The BHC mechanism of course presupposes that
the stellar-mass black holes have had sufficient time since the formation
of the galaxy to segregate into a central cusp. The segregation occurs by
dynamical friction on the timescale $t_d \approx \frac{m_s}{m_b} t_r$, where $t_r$ is the
relaxation timescale \e t_r(r)=0.34 \frac{\sigma(r)^3}{G^2 m_s \rho(r) \ln
\Lambda}
\label{teq}
\st (Binney \& Tremaine 1987), $m_b \sim 7\msun$ is the mass of a typical segregating black hole and $\ln
\Lambda$ is the Coulomb logarithm which we take as $\approx 10$. Using the
expressions for density and velocity dispersion above, it is easily shown
that the relaxation and dynamical friction timescales have a constant minimum value below the radius
of influence and rise as $r^{2}$ outside it. Thus we can calculate the
minimum time required to form a cluster of black holes from the minimum dynamical friction timescale at the radius of influence $t_d(r_i)$. For the MW, the minimum dynamical friction
timescale is $t_M\approx 1.4$Gyr (Miralda-Escud{\'e} \& Gould 2000) which
implies that a black hole cusp would form within a small fraction of the age of our
galaxy and produce HVSs. Equation (\ref{teq}) implies that the minimum dynamical friction timescale in M31 is \e t_A=t_M
\left(\frac{\sigma_A}{\sigma_M}\right)^3 \left(\frac{\rho_M}{\rho_A}\right)
\st which gives $t_A\approx340 t_M\approx480$Gyr. This greatly exceeds the
age of the universe, and it follows that no central black hole cluster has
formed in M31\footnote{A caveat to this conclusion involves violent
relaxation (Lynden-Bell 1967) which could result from a merger
event. However, the effect of a merger is expected to be episodic and would
not lead to a cumulative segregation process that is considered here over
long periods of time.}. We thus do not expect any HVS production in M31 by the BHC mechanism, and
we will not consider this mechanism in our calculations.

We now consider the effect of the increased MBH mass and different stellar
environment in M31 on the TB mechanism. In the TB mechanism, a binary which
approaches the MBH to within the tidal radius $r_t$ is disrupted; one of
the stars is captured and its binary companion is ejected at high
velocity. A binary will be disrupted within an orbital period $t_p$ if its
orbital specific angular momentum (per unit mass) $j$ is sufficiently
small, $j \leq j_{min} \approx \sqrt{GM_\bullet r_t} \approx \sqrt{G
M_\bullet a (M_\bullet/m_s)^{1/3}}$, where $a$ is the binary separation. This range corresponds to a
region in the angular momentum -- energy phase space which is known as the
``loss-cone''. The rate of binary disruption is either limited by the
orbital period (``full loss-cone'') or by the time for binaries to diffuse into the loss cone (``empty loss cone''); to calculate the total rate one must consider both these regimes (see Frank \& Rees 1976, Lightman \& Shapiro 1977, Cohn \& Kulsrud 1978, Magorrian \& Tremaine 1999, Yu \& Tremaine 2003 and
Perets et al.\ 2007). The cross-over from the empty to full loss-cone regimes occurs at a critical radius $r_c$ where the root-mean-square angular momentum change $\Delta j$ due to stellar encounters during one orbital period $t_p$
is equal to $j_{min}$. We can calculate the critical radius as follows. $r_c$ is defined by \e
\Delta j (r_c)= j_{min}. \st The root mean square specific angular momentum
transfer over one orbital period is (Lightman \& Shapiro 1977) \e \Delta j \approx \left(\frac{t_p}{t_r}\right)^{1/2} j_{circ} ,\st
where $j_{circ}(r)\approx \sigma(r) r$ is the specific angular momentum of
a star on a circular orbit.  This implies \e
\left(\frac{j_{min}}{j_{circ}(r_c)}\right)^2\approx\frac{t_p(r_c)}{t_r(r_c)} .\st
Noting that within the radius of influence $\sigma$ is of order the
circular velocity and that outside the radius of influence it is a constant, we
take $t_p\approx2 \pi r/\sigma(r)$. Using the expressions given previously
for $j_{min}$, $j_{circ}$ and $t_r$, we obtain an equation involving $r_c$ \e \frac{G M_\bullet a
(M_\bullet/m_s)^{1/3}}{\sigma^2(r_c) r_c^2}\approx\frac{2 \pi r_c G^2 m_s
\rho(r_c)}{0.034 \sigma^4}. \st Solving this
equation using $a=0.1$AU leads to a critical radius significantly larger than the radius of influence for both M31 and the MW ($r_i/r_c\ll 1$).

We now make an order of magnitude estimate of the HVS production rates using our density profile. As is discussed in Perets et al.\ (2007) \S4, the contribution to the total rate $\Gamma$ is $\dd \Gamma/ \dd r \propto r^{1/2}$ for $r<r_i$, $\dd \Gamma/ \dd r \propto r^{-2}$ for $r_i<r<r_c$ and $\dd \Gamma/ \dd r \propto r^{-3}$ for $r>r_c$. The continuity of $\dd \Gamma/ \dd r$ allows us to estimate the relative contributions of these three different regions to the total rate; it is easily shown that the contribution from the full loss-cone region $r>r_c$ is smaller than that from the empty loss cone region $r<r_c$ by a factor $r_i/r_c$ and is thus negligible. Perets et al.\ (2007) give an approximate expression for $\dd \Gamma/ \dd r$ in the empty loss cone regime (their Eq.\ 17):\e \frac{\dd \Gamma_e}{\dd r}\sim\frac{N_s(<r)}{r t_r(r_c)} \st where $N_s(<r)$ is the number of binaries enclosed within $r$. The total rate can thus be obtained from an integration over $r$:
\e
\Gamma\sim \int_0^{r_c} \frac{\dd \Gamma_e} { \dd r}  \dd r \sim \int_0^{\infty} \frac{\dd \Gamma_e} { \dd r}  \dd r \sim\frac{40 \pi G^2 \eta \rho_i^2 r_i^3}{0.034 \times 6 \sigma_i^3}
\label{rateeq}
\st
Equation (\ref{rateeq}) allows us to find the expected ratio of HVS
production in the MW and M31 based on the observational data mentioned
above\footnote{We also verified that this simple formula gives a rate very similar to the more exact Yu \& Tremaine 2003 calculation for the MW.}: \e
\frac{\Gamma_A}{\Gamma_M}\sim\left(\frac{\rho_A}{\rho_M}\right)^2
\left(\frac{r_A}{r_M}\right)^3
\left(\frac{\sigma_M}{\sigma_A}\right)^3\sim 0.2 \st This rough estimate
indicates that the TB mechanism produces $1$--$10$ times fewer HVSs in M31
than in the MW.

The IBH mechanism is also affected by the differences between the MW and
M31. The HVS ejection rate was calculated for M31 by Lu et al.\ (2007) by
adopting the method of Yu \& Tremaine (2003); assuming an IBH/MBH mass ratio of $\nu=0.01$ and an IBH-MBH axis of $5.9$mpc the rate is
approximately an order of magnitude lower than in the MW (of order
$10^{-5}$/yr).

To summarise, the BHC mechanism produces no HVSs in M31, and both the TB
and IBH HVS production rates are suppressed in M31 by a factor of order
$1$--$10$ compared to the MW. However, this may be partially compensated for by the somewhat higher ejected star velocities in M31 (Yu \& Tremaine 2003) which decrease the proportion of bound ejected stars. The number of HVSs produced over the past
$10^{10}$ years in the MW is expected to be somewhat above $10^6$. For our
calculations with both these mechanisms we thus assume that the total
number of HVSs produced in M31 is $10^6$. This estimate is fairly
uncertain and relies on the assumptions that the stars are isotropically distributed and that the effects of any massive perturbers (Perets et al.\ 2007) are equal in the MW and in M31. The number of HVSs produced could easily be an order of
magnitude lower or higher than our estimate. However, our statistical results can be simply
scaled to a different number of ejected stars based on future observations and calculations.

\subsection{HVS Velocities and Masses}
We now consider the probability distributions for the HVSs' ejection velocities and
masses. Both the TB and IBH models have the welcome feature that the form
of the probability distribution for the velocity of an ejected star is at
most weakly dependent on its mass: for the IBH mechanism, the velocity is
entirely independent of the mass of the HVS (as the HVS acts as a test
particle in the gravitational field of the much heavier black holes); for the TB
mechanism, the velocity of a HVS of mass $m_1$ shows a very weak dependence
$\propto \sqrt{m_2} (m_1+m_2)^{-1/6}$, where $m_2$ is the mass of the other
binary component (Hills 1988). We thus approximate the velocity and mass distributions as independent, which makes our
numerical exploration of their statistics much simpler.

For the TB model, Sesana et al. (2007a) apply the results of Bromley et
al. (2006) to binaries with a semi-major axis $a$ distributed uniformly in
$\ln a$ (Heacox 1998). Following their Figure 1 (TBf model), the TB
mechanism gives rise to a distribution of star ejection velocities $v$
(after escaping the gravitational field of the MBH) of approximately \e
P(v) \propto \left\{\begin{array}{cl} v^{-4.9}, & v \geq v_l \\ 0, & v<v_l
\end{array}\right.
\label{vh}
\st where $P(v) \dd v$ provides the probability that a star has an ejection
speed in the interval $(v, v+dv$). While this velocity distribution was derived for the MW, we will assume that the power
law index of the distribution is characteristic of the HVS production
mechanism and does not depend strongly on the MBH mass, so that
this distribution can also be applied to M31. $v_l$ is the lower cutoff velocity,
which we define as the maximum speed with which stars do not escape the M31
halo over $10^{10}$ years for any ejection direction. $v_l$ is thus very
similar to (but slightly smaller than) the escape velocity.

The HVS velocity distribution for the IBH model was calculated numerically
for the MW by Sesana et al.\ (2006, 2007a,b). With an IBH/MBH mass ratio of
$\nu=1/729$ and an initial orbital eccentricity of $e=0.9$, they consider
the entire in-spiral of the black hole and obtain a HVS velocity
distribution $\propto v^{-2.5}$. This distribution, with the high velocity tail, is nearly independent of the assumed values of $\nu$ and $e$ (A. Sesana, private communication). Again, we assume that the M31 HVS
ejection velocity distribution has the same power law index as was found in
the MW and thus obtain for M31 \e P(v)\propto \left\{\begin{array}{cl}
v^{-2.5}, & v \geq v_l \\ 0, & v<v_l \end{array}\right.
\label{vi}
\st
As explained previously, the initial speed $v_0$ we use for our simulations of HVS trajectories is given by $v_0=\sqrt{v^2-v_{bd}^2}$ (taking into account the slowing by the M31 bulge and disc.)

The dependence of HVS ejection probability on stellar mass is very weak for both the TB and IBH mechanisms, and so we will assume that the distribution of ejected stellar masses follows the distribution of main sequence stars in the galactic centre. However, the
star formation history in M31 over the past $10^{10}$ years is not known,
and so for simplicity we will assume a Salpeter initial mass function
$\frac{\dd N}{\dd m}\propto m^{-2.35}$. As observations indicate that the formation of low mass stars is suppressed in galactic centres (Maness et al.\ 2007) and very low mass stars are difficult to observe, we cut off the mass distribution below $1\msun$. However, since the lifetime of $1\msun$ stars is roughly the same as the Local Group age, the distribution of the longer lived lower mass stars can be obtained by simply rescaling the $1\msun$ distribution to reflect any differences in the total number of stars.  As discussed previously, taking stellar
lifetime into consideration implies that the HVSs ejected from M31 thus have
masses distributed according to \e P(m)\propto \left\{\begin{array}{cl}
m^{-4.85}, & m \geq 1 \msun \\ 0, & m< 1 \msun
\end{array}\right.
\label{m}
\st
We assume that the HVSs are ejected isotropically from the M31 centre.

\section{Numerical Methods and Analysis of our Model}
\subsection{Numerical Methods}
In order to obtain HVS distributions and velocities we must simulate a
large number of individual stellar trajectories. The basic challenge of
this calculation is to integrate Newton's gravitational force law. The
integration method we use is the Verlet method (Verlet 1967), which has a
number of advantages: it is symmetric (i.e. it uses information from both
ends of a step), it is a fourth order method, and it is sufficiently fast
to allow a large number of trajectories to be calculated; many HVS
trajectories are needed to obtain statistically valid information. A fixed
step size is used for our integration as this greatly simplifies the
determination of the HVS distributions.

To find one HVS trajectory, our integrator was run for a time corresponding
to the age of the Local Group, which we take to be $10^{10}$ years. We
tested our program by comparing its results with a number of analytically
calculated trajectories for different force laws. Our numerical results
agreed with the analytical solutions for all the configurations we tested
(e.g.\ Keplerian orbits, simple harmonic motion). For our numerical
simulations, we chose to use $10^4$ integration steps per trajectory. This number of steps
was not computationally expensive and gave results which agreed very well with analytic solutions. It was verified that the trajectories were identical for a wide
range of step-numbers (from $500$ to $2\times10^5$), which suggests that
the integration scheme is stable and that our results are reliable.

\subsection{Symmetry Considerations}
Our model for the Local Group has a number of important symmetries, which
can be exploited to make the necessary numerical calculations more
efficient. As is seen in Figure \ref{localgroup}, the system exhibits
rotational symmetry along the $z$ axis connecting the two galaxies, which
means that every trajectory of an HVS from the centre of M31 follows a path
which lies in a single plane through the inter-galactic axis. This feature
allows us to reduce our problem from three to two dimensions (and thereby increase the computational efficiency), by
``folding'' all possible trajectories into one cross-section plane within
which we perform all our calculations. To regain
three dimensional results we can simply deproject our values. An isotropic ejection in
three dimensions corresponds to the following distribution in two
dimensions: \e P(\theta)=\frac{1}{2}\sin \theta
\label{angle}
\st where $\theta$ is the angle between the intergalactic $z$-axis and the
initial HVS velocity vector.

In addition, the reflection symmetry of the upper and lower half planes
means that a ``reflected'' trajectory is also a valid one. We can thus
limit ourselves to sampling trajectories that begin in the upper half
plane.

The final and most important symmetry involves time. In our model the mass
distribution as shown in Figure \ref{localgroup} is static. This means that
every point (not just the endpoint) of a trajectory is a
possible current location of a HVS. For every simulated trajectory we thus
obtain not one, but many possible current star positions.

The choice of a two-dimensional static Local Group model over a three-dimensional dynamic model reduces the computational time for the HVS distributions by a factor of $10^{7}$.

\subsection{Calculation of the HVS Distributions}
To find the HVS distributions and velocities, we must calculate a large
number of HVS trajectories. The initial conditions for each trajectory are
determined by sampling randomly from the distributions for the HVS ejection
angle (Eq.\ \ref{angle}) and speed (Eq.\ \ref{vh}/\ref{vi}). We use Park
and Miller's (1988) ``Minimal Standard'' random number generator together
with the transformation method detailed in Press et al.\ (1996) to obtain
initial conditions with the required distribution. $1.2 \times 10^6$
trajectories were thus calculated, which correspond to 1100 independent
samples in both angle and velocity. For each trajectory we used $10^4$
timesteps. This means that we obtain a total of $1.2\times 10^{10}$ points,
each of which is a possible current position of a HVS. However, due to the
finite stellar main-sequence lifetime, each of these points does not carry
equal weight when determining the distribution of HVSs within the Local
Group.

We now consider how to obtain the distribution of HVSs from the raw
trajectory data. The lifetime $t_l$ of a star of mass $m$ is assumed to be
$t_l=10$Gyr$(m/\msun)^{-2.5}$ (Kippenhahn \& Weigert 1990). We take the age
of the Local Group to be $t_{\mathrm{age}}=10^{10}$yr. Every point on a
trajectory corresponds to an HVS travel time $t_t$ which can range from
zero to $t_{\mathrm{age}}$. Before being ejected from the centre of M31, a
star is assumed to have resided in the galactic centre for a ``dwell time''
$t_d$. For a star with mass $m$ and corresponding main-sequence lifetime $t_l$, a point
along a trajectory corresponding to a travel time $t_t$ can contain a ``live''
star if $t_t+t_d \leq t_l$. Therefore the probability $F(t_t)$ of a star at a point corresponding to $t_t$ being ``live'' is equal to the probability
that $t_d<(t_l-t_t)$.

To determine this probability we must consider the distribution of dwell
times $P(t_d)$. There are two conditions on $t_d$. A ``live'' star cannot
have resided in the galactic centre for more than its main-sequence
lifetime $t_l$, so that $t_d<t_l$. Of course, it can also not have been
there before the formation of the Local Group $t_{\mathrm{age}}=10^{10}$ years
ago, which implies $t_d+t_t<t_{\mathrm{age}}$. Thus the maximum value of
$t_d$ is $t_{d,\mathrm{max}}=$min$(t_l,t_{\mathrm{age}}-t_t)$. We assume
that a live star is equally likely to have been in the galactic centre for
any time up to the maximum permissible value $t_{d,\mathrm{max}}$, and the
dwell time thus obeys a uniform probability distribution
$P(t_d)=1/t_{d,\mathrm{max}}$ between $0$ and $t_{d,\mathrm{max}}$.

The probability $F(t_t)$ of a star at a point with a travel
time $t_t$ being live is equal to the likelihood of fulfilling the condition
$t_d<(t_l-t_t)$, i.e.  \e F(t_t)=\int_0^{t_l-t_t} P(t_d) \dd t_d \st

The different ranges of this integral imply different forms for $P(t_d)$,
which complicates this (otherwise straightforward) integral. 
If $t_l<0.5 t_{\mathrm{age}}$
\e
F(t_t)=\left\{\begin{array}{cl}
	 1-\frac{t_t}{t_l}, & t_t<t_l \\
	 0, & t_t>t_l
	   \end{array}\right.
\st

If $t_l>0.5 t_{\mathrm{age}}$
\e
F(t_t)=\left\{\begin{array}{cl}
	 1-\frac{t_t}{t_l}, & t_t<t_{\mathrm{age}}-t_l \\
	 1-\frac{t_{\mathrm{age}}-t_l}{t_{\mathrm{age}}-t_t}, & t_l>t_t>t_{\mathrm{age}}-t_l \\
	 0, & t_t>t_l
	   \end{array}\right.
\st

The density of HVSs within the Local Group is thus calculated as
follows. We partition the ejected stars into six mass bins containing equal
numbers of HVSs according to the mass distribution given by equation
(\ref{m}). The mass ranges corresponding to these bins are $(1-1.05)\msun$,
$(1.05-1.11)\msun$, $(1.11-1.20)\msun$, $(1.20-1.33)\msun$,
$(1.33-1.59)\msun$ and $(>1.59)\msun$ for mass bins one through six. The
average mass in each bin is used to determine the corresponding $t_l$,
which allows the calculation of $F(t_t)$ for all values of $t_t$. We then
divide the upper half plane of Figure \ref{localgroup} into an array of
($300\times 600$) cells and add up the number of points in each cell, with
each point weighted by $F(t_t)$. Dividing by the total number of trajectory
points ($1.2\times 10^{6}\times10^4$) as well as the volume corresponding to each
cell and multiplying by the total number of HVSs produced gives a number
density $n(d,z)$ of stars/kpc$^3$ for each mass bin; the coordinates ($d$, $z$) are defined in
Figure \ref{localgroup}.

In addition to the distributions for the six mass bins, we also consider
massive stars with masses above $3 \msun$. These are important
observationally due to their high luminosity (the observed MW HVSs have mainly been $3-4\msun$ stars). We take their lifetime to be represented by $t_l=0.35$Gyr (Schaller et al.\ 1992).


\section{Results}
\subsection{Star Trajectories}
We first investigated the shapes and characteristics of individual star
trajectories. Single HVSs were expelled from the centre of M31 with a
variety of initial conditions and their subsequent paths were found by
integration of the gravitational force. For large ejection velocities
($v>900$$\ \mathrm{km} \ \mathrm{s}^{-1}$) the stars escaped the Local Group
entirely, without being deflected significantly from their original
straight trajectories. For velocities smaller than the lower cutoff velocity, the
stars were bound to M31 and precessed about its centre in loop orbits. In
addition to the bound and expelled stars, stars released under certain
angles with a small range of velocities $\approx 650-900$$\ \mathrm{km} \
\mathrm{s}^{-1}$ were gravitationally pulled into the MW or towards the intergalactic axis. Their
trajectories had very low average speeds of approximately
$100-200$$\ \mathrm{km} \ \mathrm{s}^{-1}$, as these stars have just enough energy to escape M31's attraction and thus spend a large
amount of time at a low velocity.

The value of the lower cutoff velocity was determined as follows. We ejected a
large number of stars out of the centre of M31 with a fixed speed but in
different directions, and observed whether or not all stars remained within the M31 halo over
the age of the Local Group. Stars
released towards the centre of the Local Group escaped M31 with lower velocities than
ones released away from the centre. By trial and error, we were able to determine that the highest speed at which all stars remained bound was
$v_l=650$$\ \mathrm{km} \ \mathrm{s}^{-1}$.

An interesting feature of the system emerged during our analysis. It was
found that the shape of the trajectories was very sensitive to the initial
conditions for some parameter values. In some cases, a change of less than
one percent in the starting velocities lead to the ejected stars following
entirely different trajectories. While the paths with slightly different
starting conditions were identical initially, it was found that the
trajectories split after a certain time. These characteristics are
indicative of chaotic behaviour in some regions of phase space. It was
found that such chaotic behaviour was most common for trajectories which
passed through the region between the two galaxies. The chaotic sensitivity
to boundary conditions appeared to require the presence of three
gravitating objects (the two galaxies and the diffuse group medium); when
only two of these objects were present, the system no longer showed chaotic
behaviour.

While it is difficult to make precise calculations of the trajectories of
individual stars in a chaotic system, our numerical experiments
demonstrated that statistical properties, such as HVS spatial and velocity
distributions, were not sensitive to slight changes in the initial
conditions.

\subsection{HVS Spatial Distributions}
To determine whether the HVSs from M31 are observable, it is necessary to
find the HVS distribution within the Local Group. By integrating $1.2
\times 10^6$ trajectories numerically and processing the points on the
trajectory as described previously, the HVS distributions were found for
both the TB and IBH mechanisms. We obtain distributions and average radial
and transverse velocities for all six mass bins and the massive stars;
Figures 2 and 3 show the results for the first and third mass
bins respectively. Distributions and velocities for the ($>3 \msun$) stars
are depicted in Figure 4.
\begin{figure*}
     \centering \subfigure[$\log_{10}$ of the number of stars per cubic
     kpc (TB)
     ]{\includegraphics[width=.48\textwidth,clip]{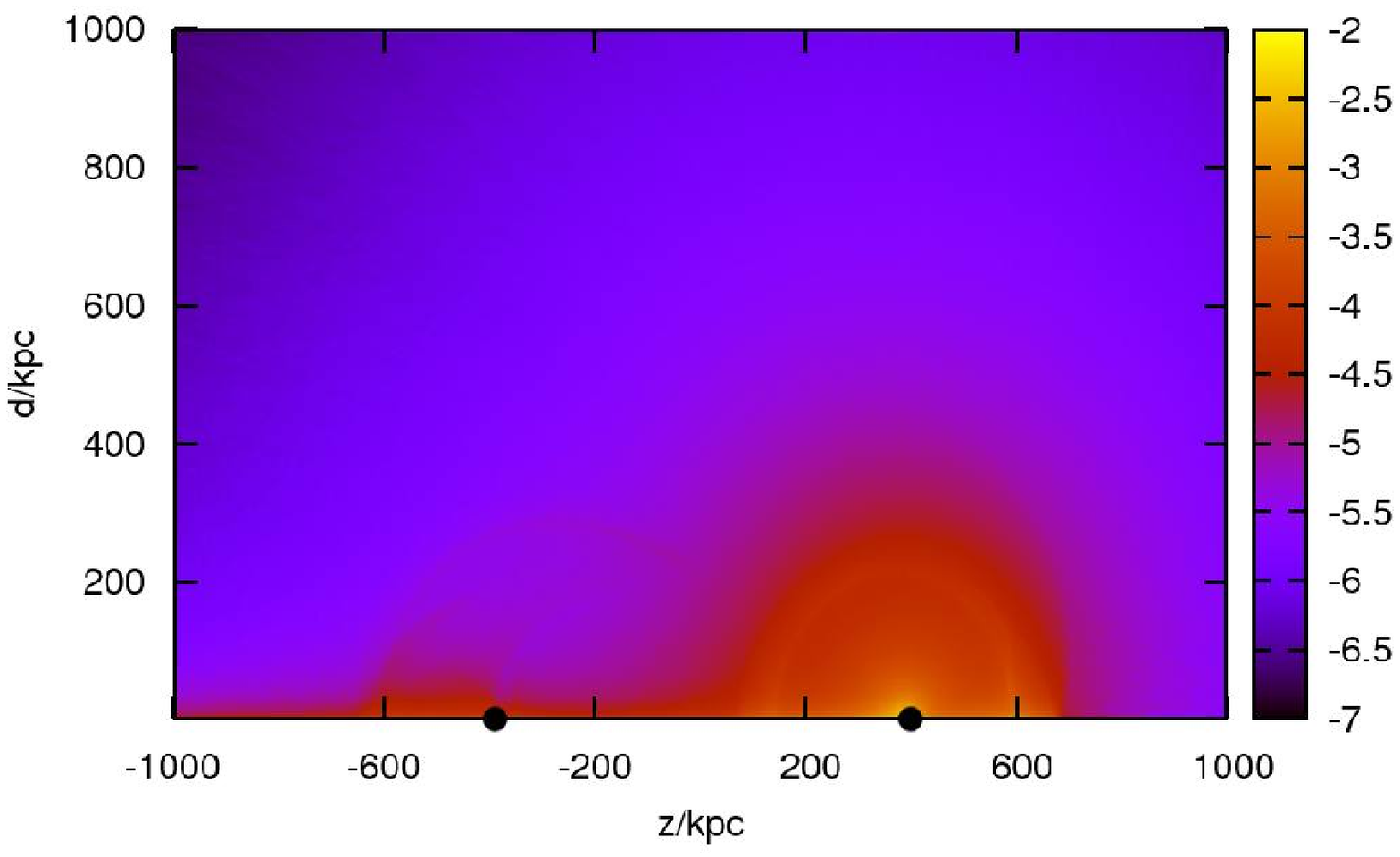}}
     \subfigure[$\log_{10}$ of the number of stars per cubic kpc (IBH)
     ]{\includegraphics[width=.48\textwidth,clip]{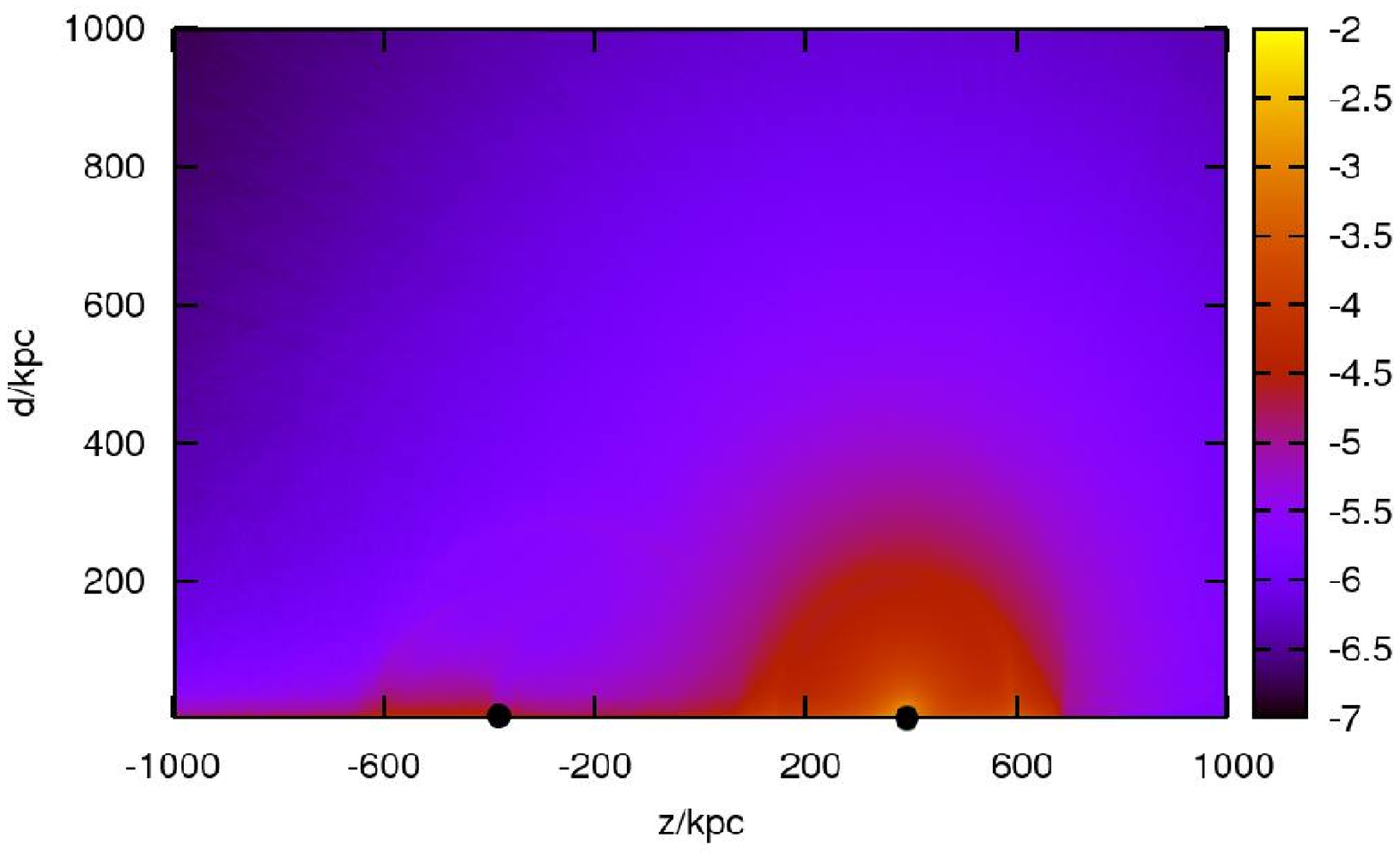}}\\
     \subfigure[average radial velocities in $\ \mathrm{km} \
     \mathrm{s}^{-1}$ (TB)
     ]{\includegraphics[width=.48\textwidth,clip]{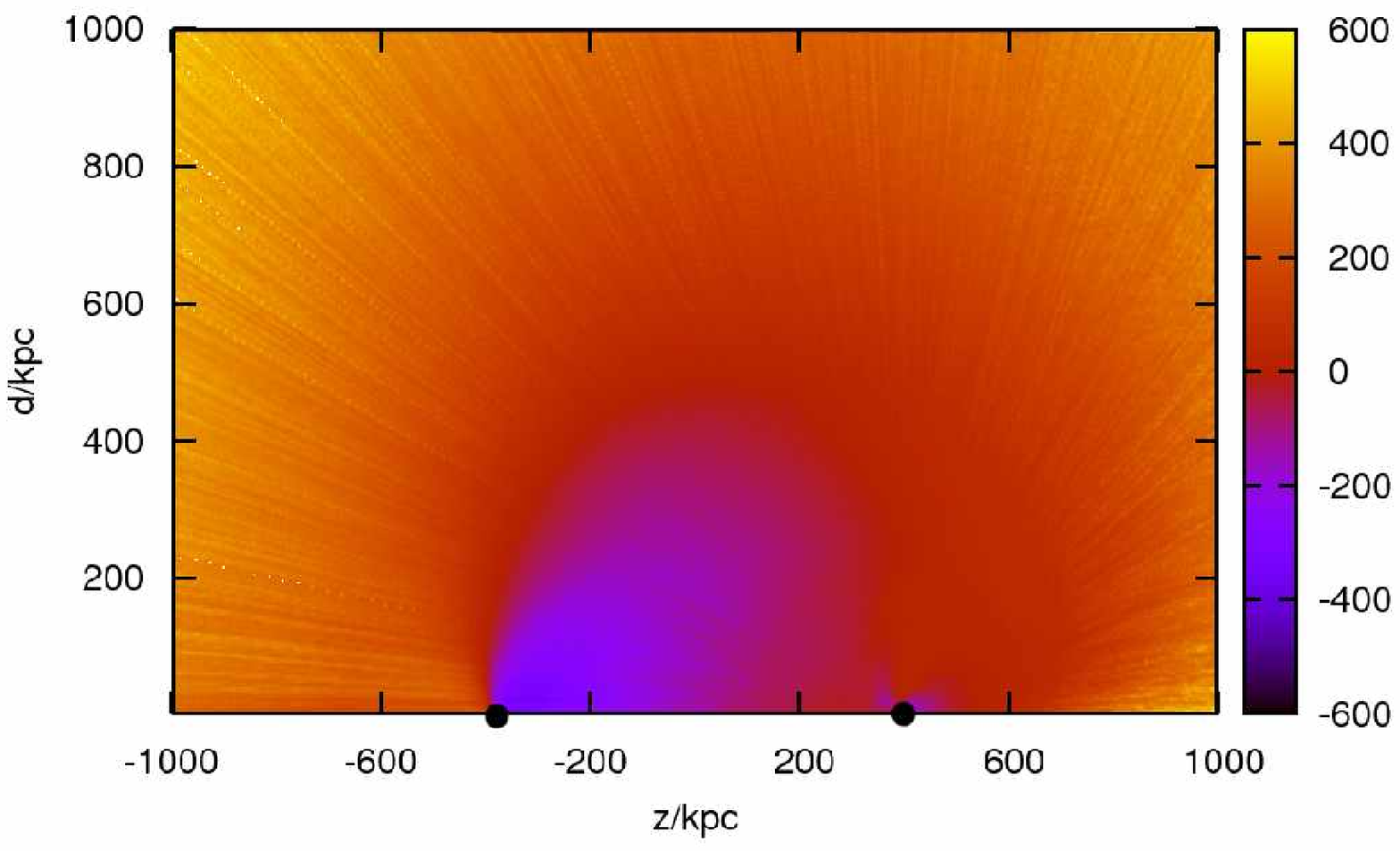}}
     \subfigure[average radial velocities in $\ \mathrm{km} \
     \mathrm{s}^{-1}$ (IBH)
     ]{\includegraphics[width=.48\textwidth,clip]{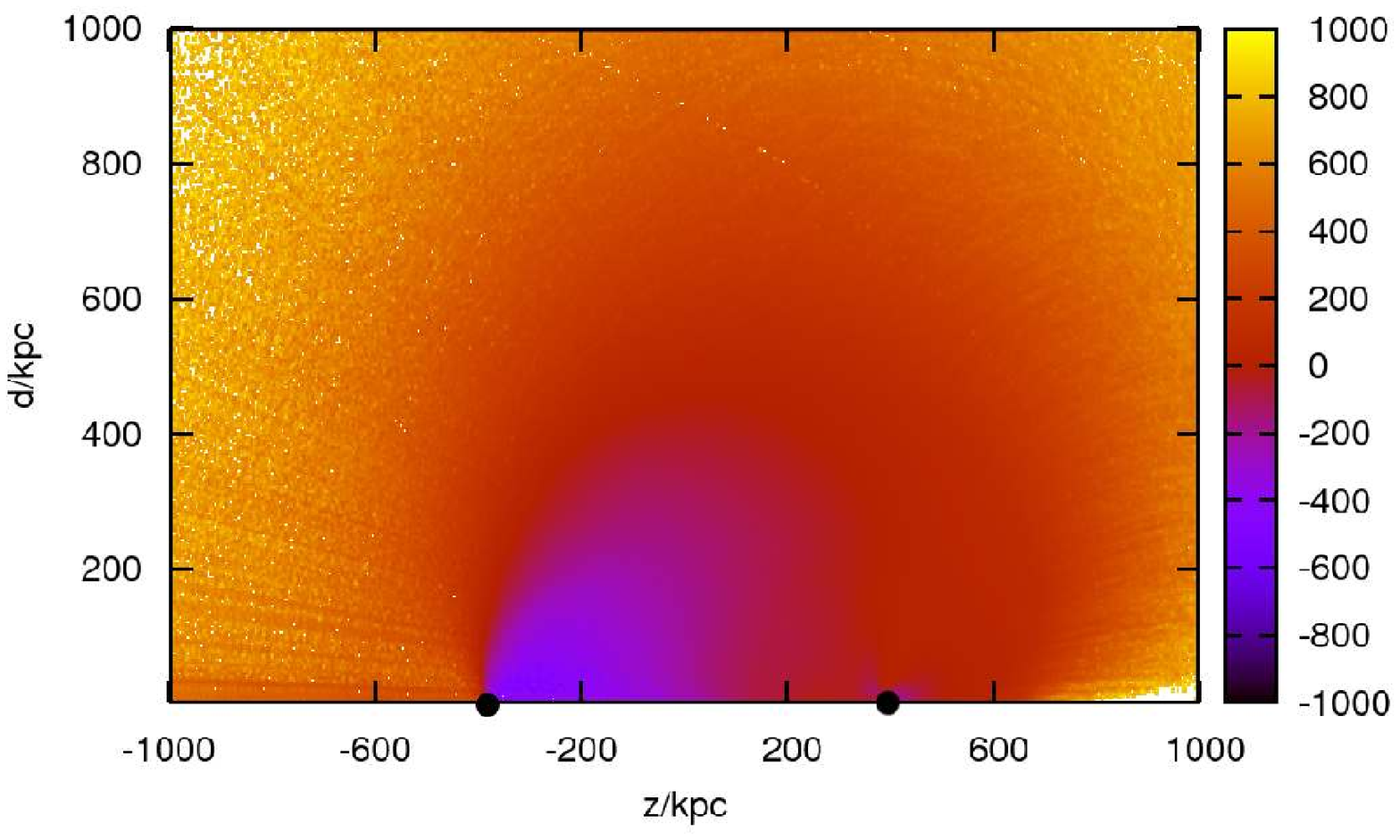}}\\
     \subfigure[average transverse velocities in $\ \mathrm{km} \
     \mathrm{s}^{-1}$ (TB)
     ]{\includegraphics[width=.48\textwidth,clip]{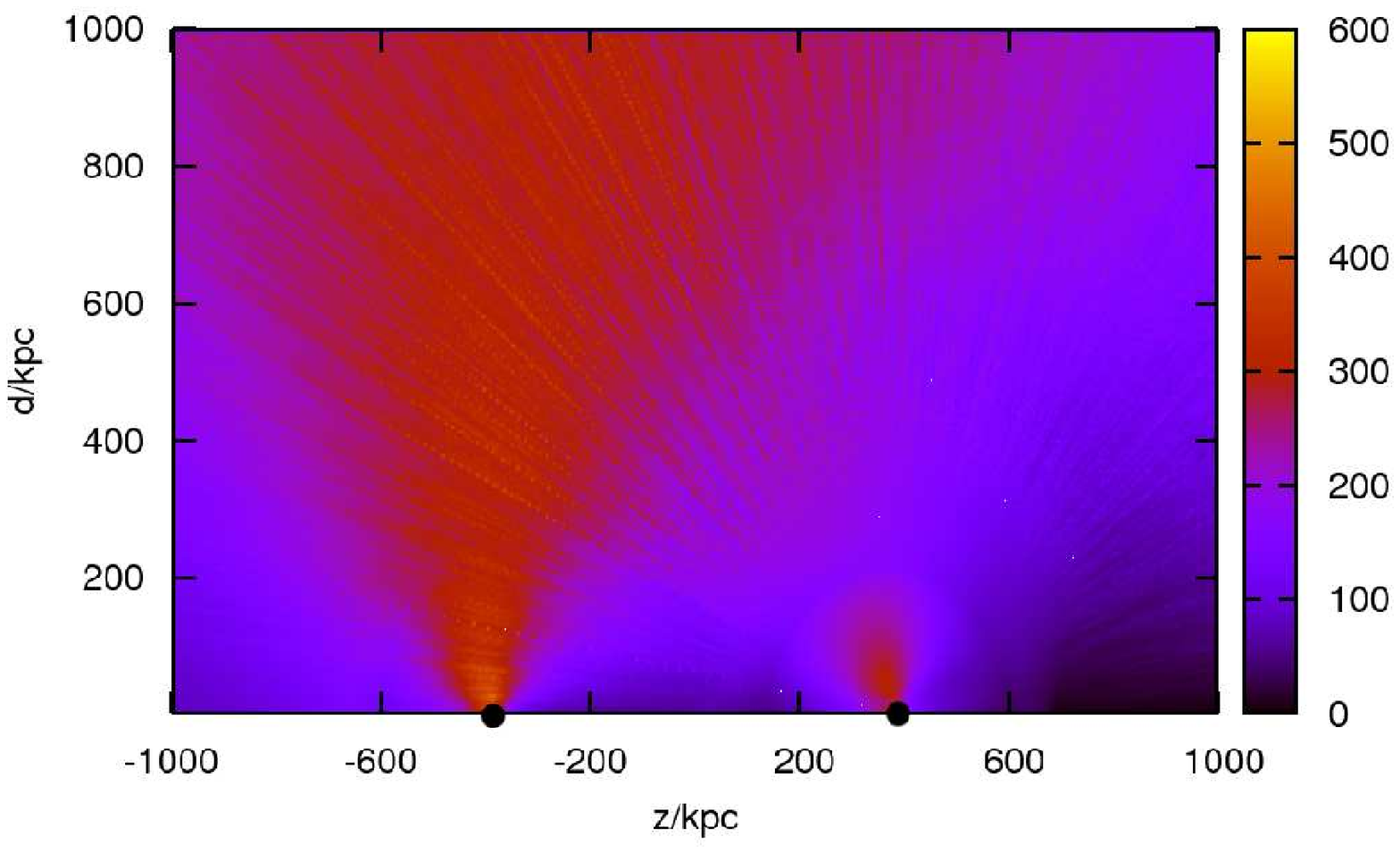}}
     \subfigure[average transverse velocities in $\ \mathrm{km} \
     \mathrm{s}^{-1}$ (IBH)
     ]{\includegraphics[width=.48\textwidth,clip]{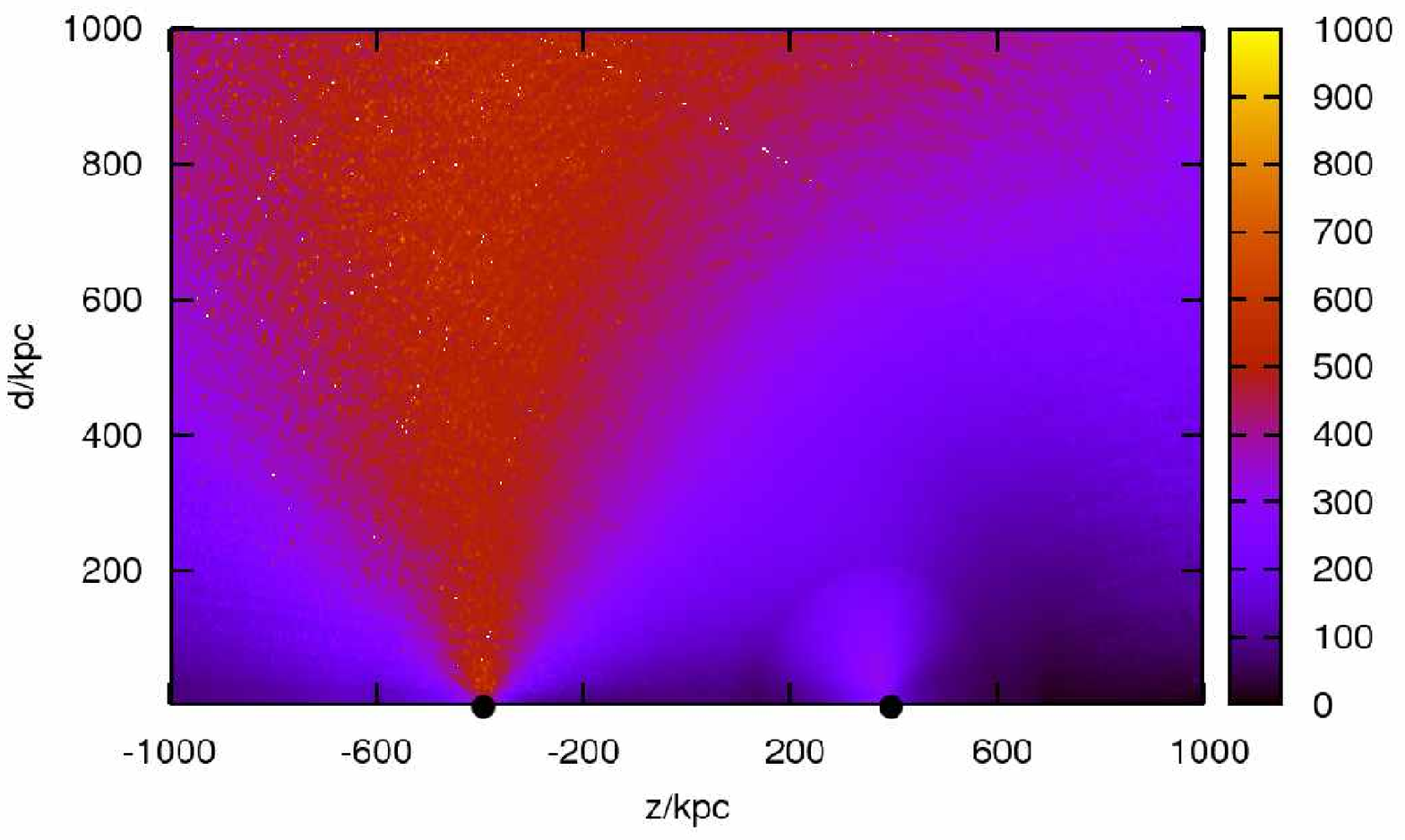}}\\
     \label{tbf} \caption{Main-sequence HVS number densities, radial velocities and
     transverse velocities (with respect to the MW galactic centre) for
     stellar masses of $(1$--$1.05) \msun$. Plots are shown for the TB
     mechanism ({\it Left column}) and the IBH mechanism ({\it Right
     column}). White colour indicates a value off the scale on the bar to
     the right of the plots. The positions of the MW and M31 centres are
     marked with black dots, with the MW on the left. See Figure 1 for the definition of the $(d,z)$ coordinates.}
\end{figure*}

\begin{figure*}
     \centering \subfigure[$\log_{10}$ of the number of stars per cubic
     kpc (TB)]{\includegraphics[width=.48\textwidth,clip]{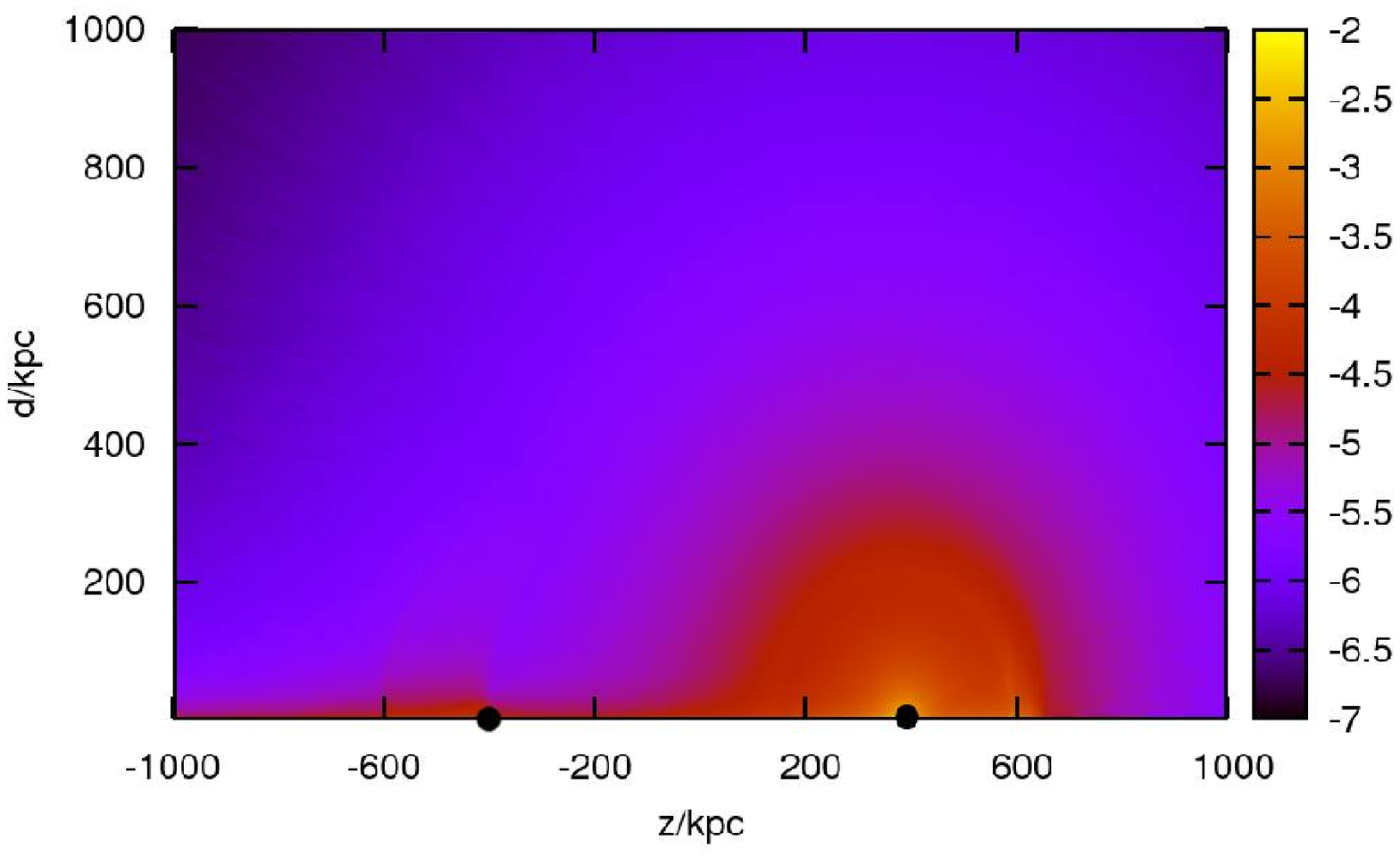}}
     \subfigure[$\log_{10}$ of the number of stars per cubic kpc (IBH)
     ]{\includegraphics[width=.48\textwidth,clip]{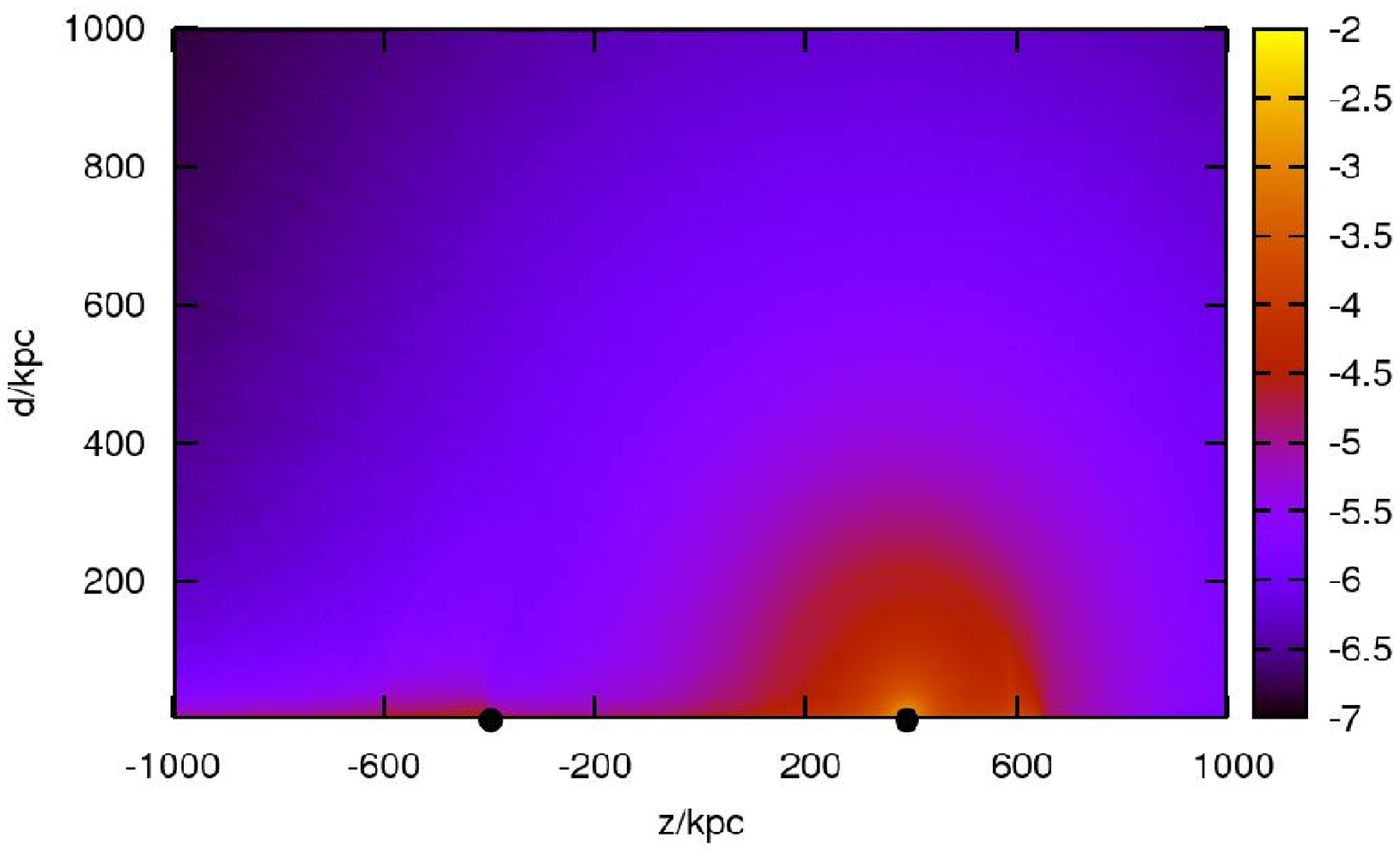}}\\
     \subfigure[average radial velocities in $\mathrm{km} \
     \mathrm{s}^{-1}$ (TB)
     ]{\includegraphics[width=.48\textwidth,clip]{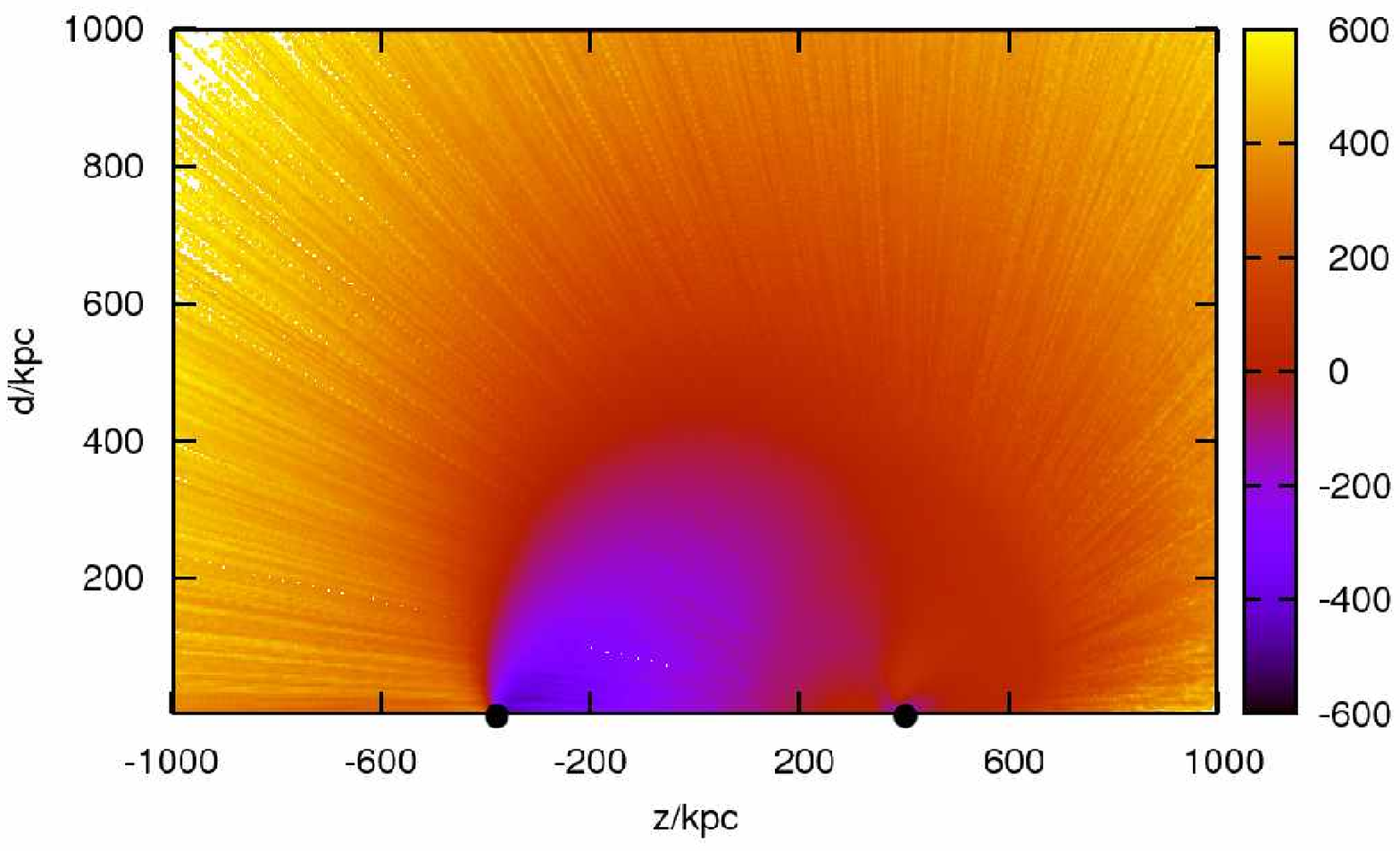}}
     \subfigure[average radial velocities in $\mathrm{km} \
     \mathrm{s}^{-1}$ (IBH)
     ]{\includegraphics[width=.48\textwidth,clip]{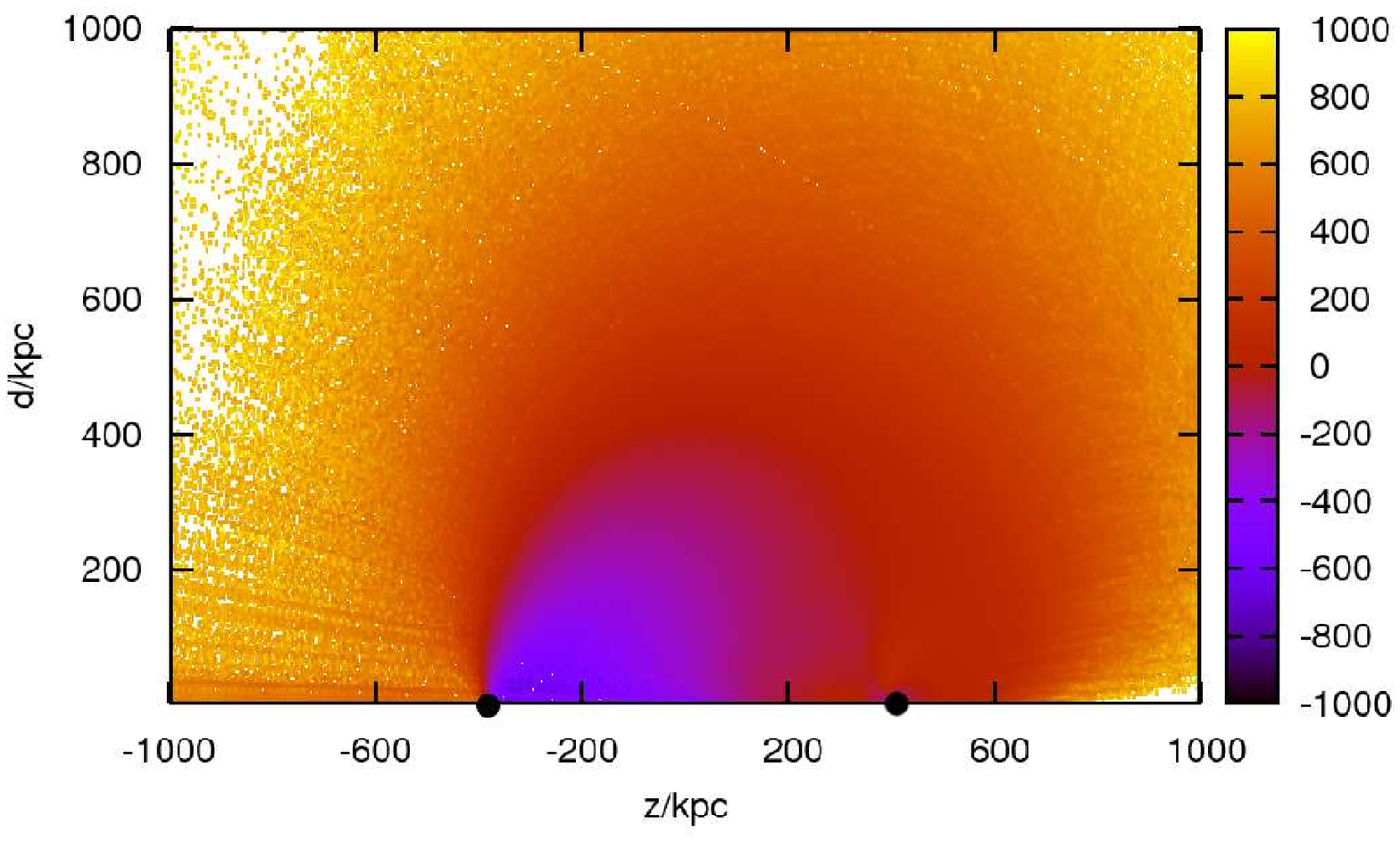}}\\
     \subfigure[average transverse velocities in $\mathrm{km} \
     \mathrm{s}^{-1}$ (TB)
     ]{\includegraphics[width=.48\textwidth,clip]{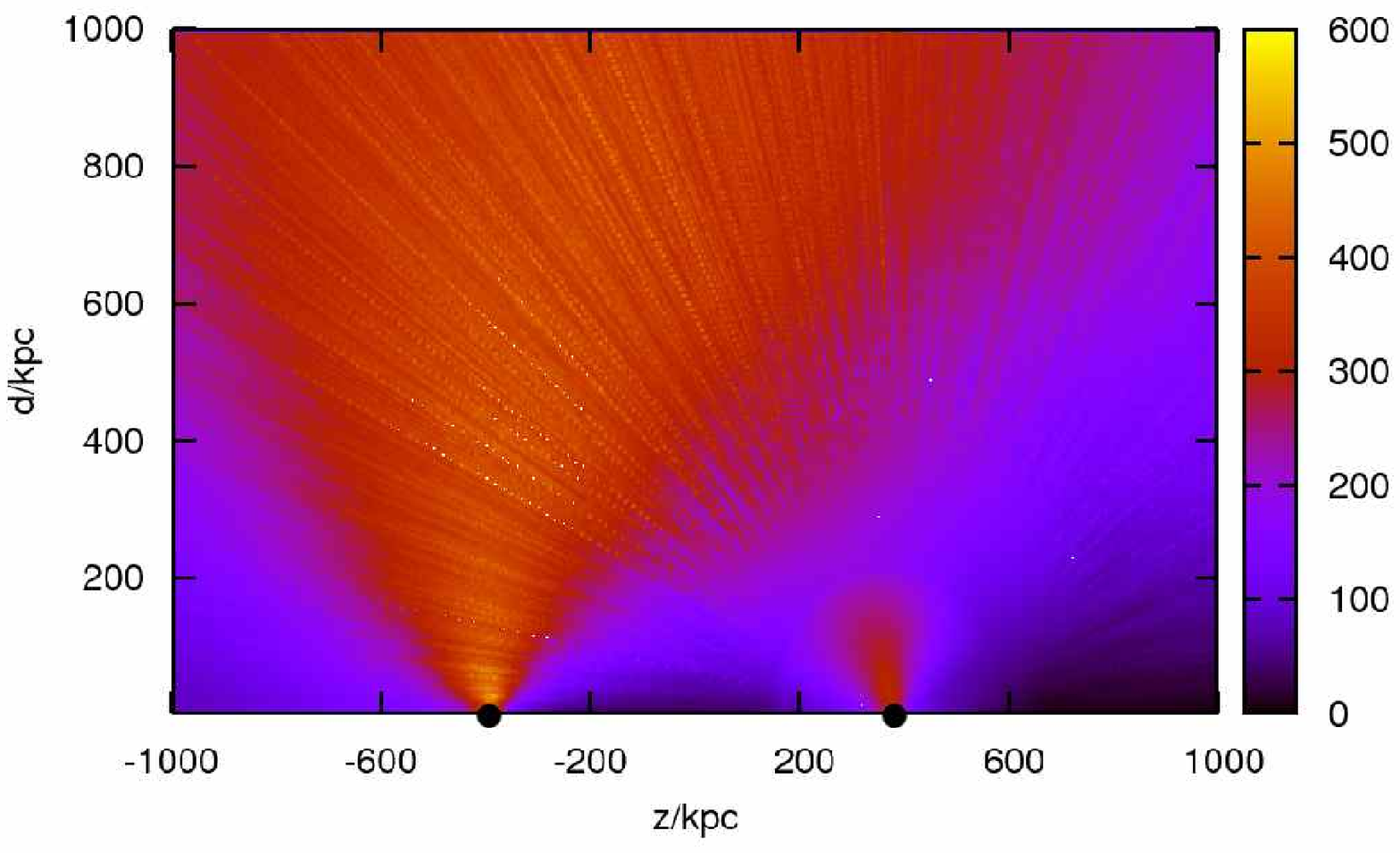}}
     \subfigure[average transverse velocities in $\mathrm{km} \
     \mathrm{s}^{-1}$ (IBH)
     ]{\includegraphics[width=.48\textwidth,clip]{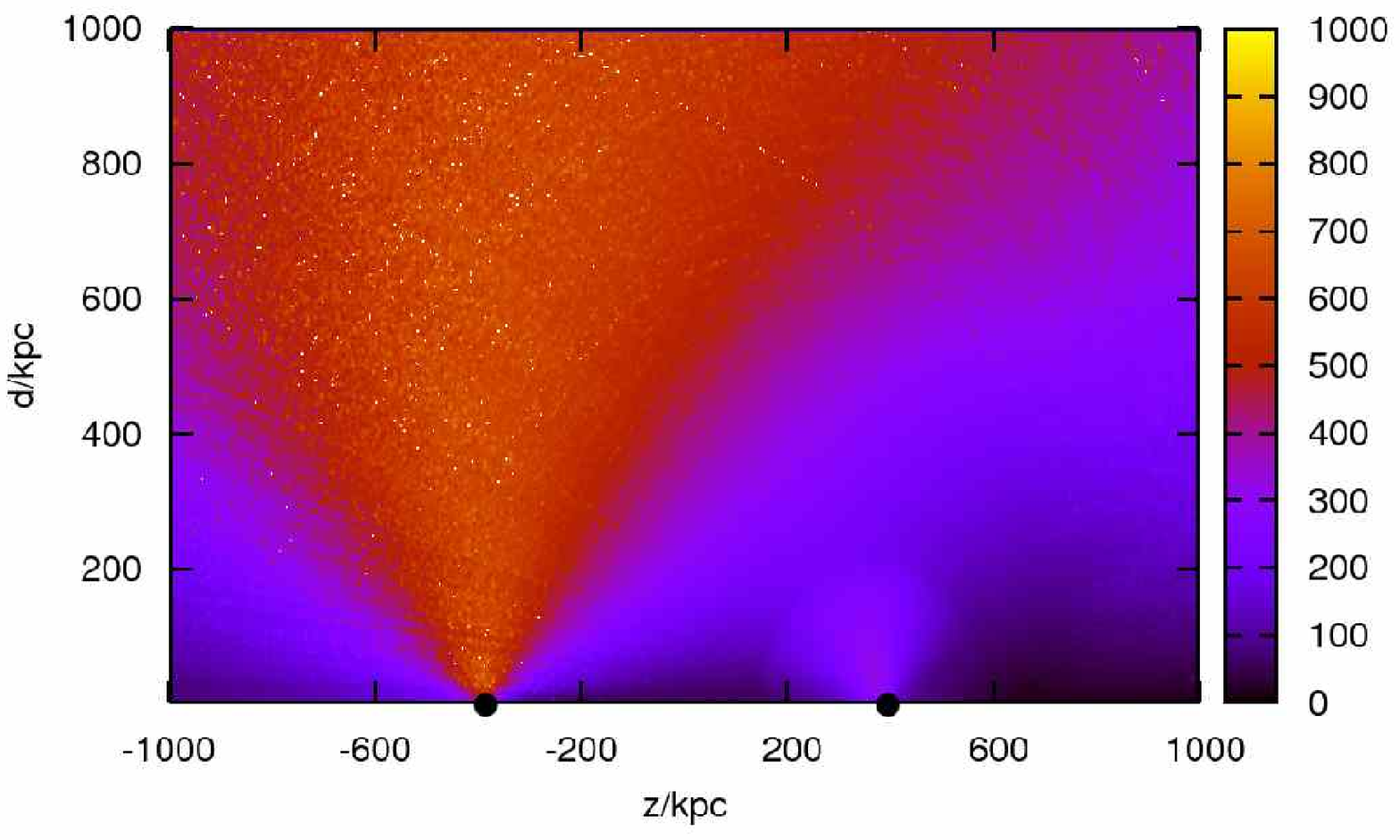}}\\
     \label{tbf} \caption{Same as Fig. 2 for
     stellar masses of $(1.11$--$1.20) \msun$.}
\end{figure*}

\begin{figure*}
     \subfigure[$\log_{10}$ of the number of stars per cubic kpc
     (TB)]{\includegraphics[width=.48\textwidth,clip]{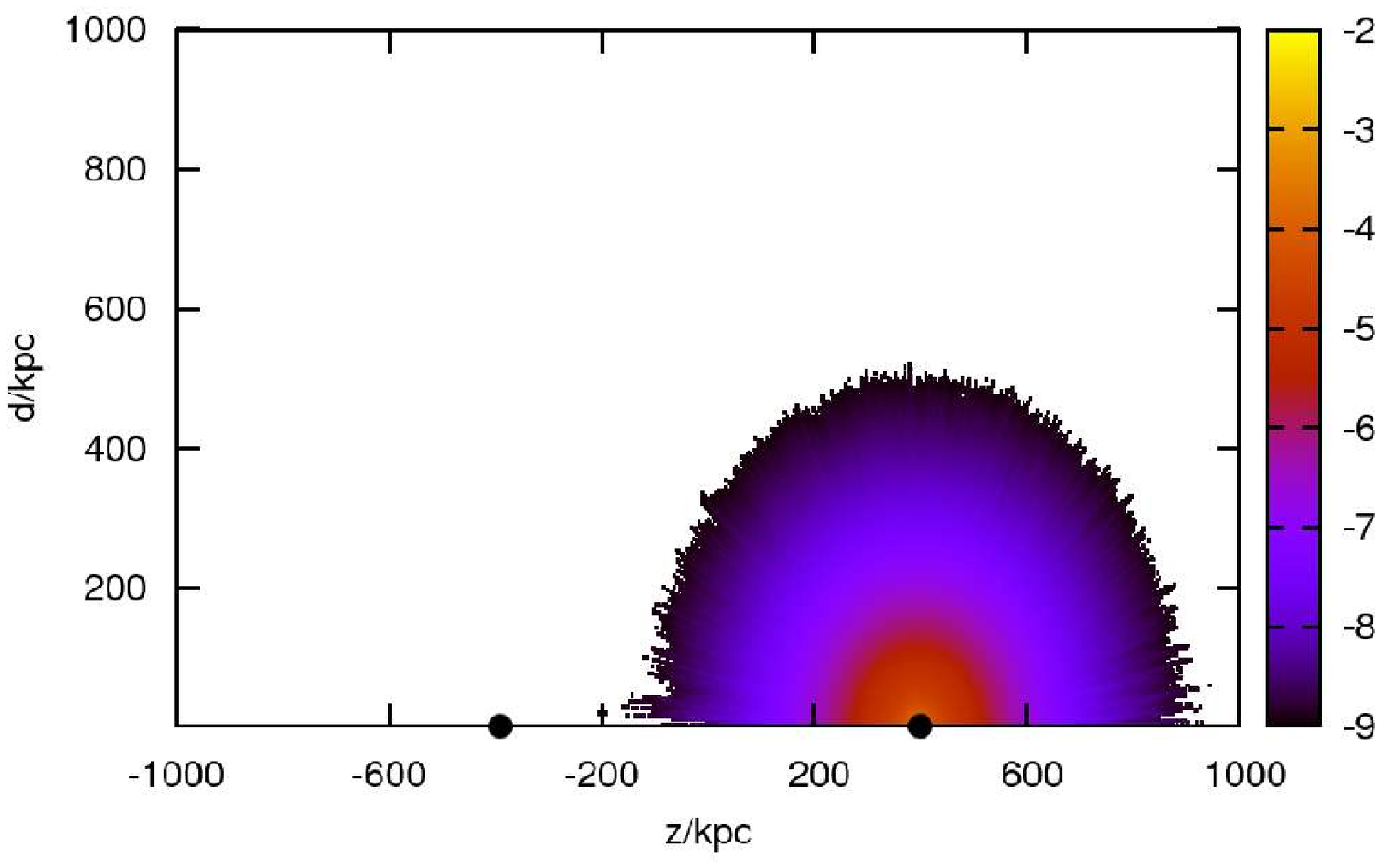}}
     \subfigure[$\log_{10}$ of the number of stars per cubic kpc (IBH)
     ]{\includegraphics[width=.48\textwidth,clip]{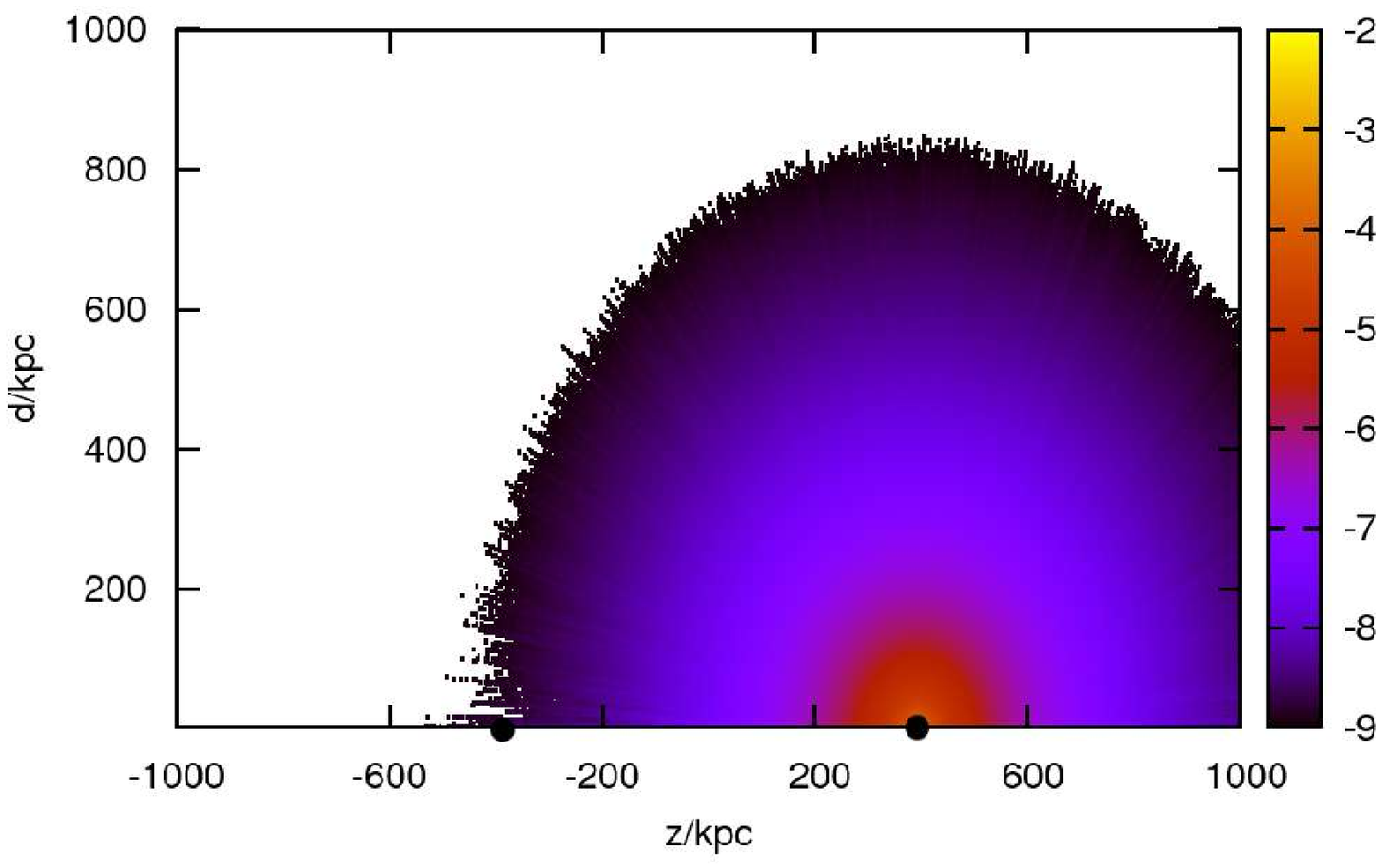}}\\
     \subfigure[average radial velocities in $\mathrm{km} \
     \mathrm{s}^{-1}$ (TB)
     ]{\includegraphics[width=.48\textwidth,clip]{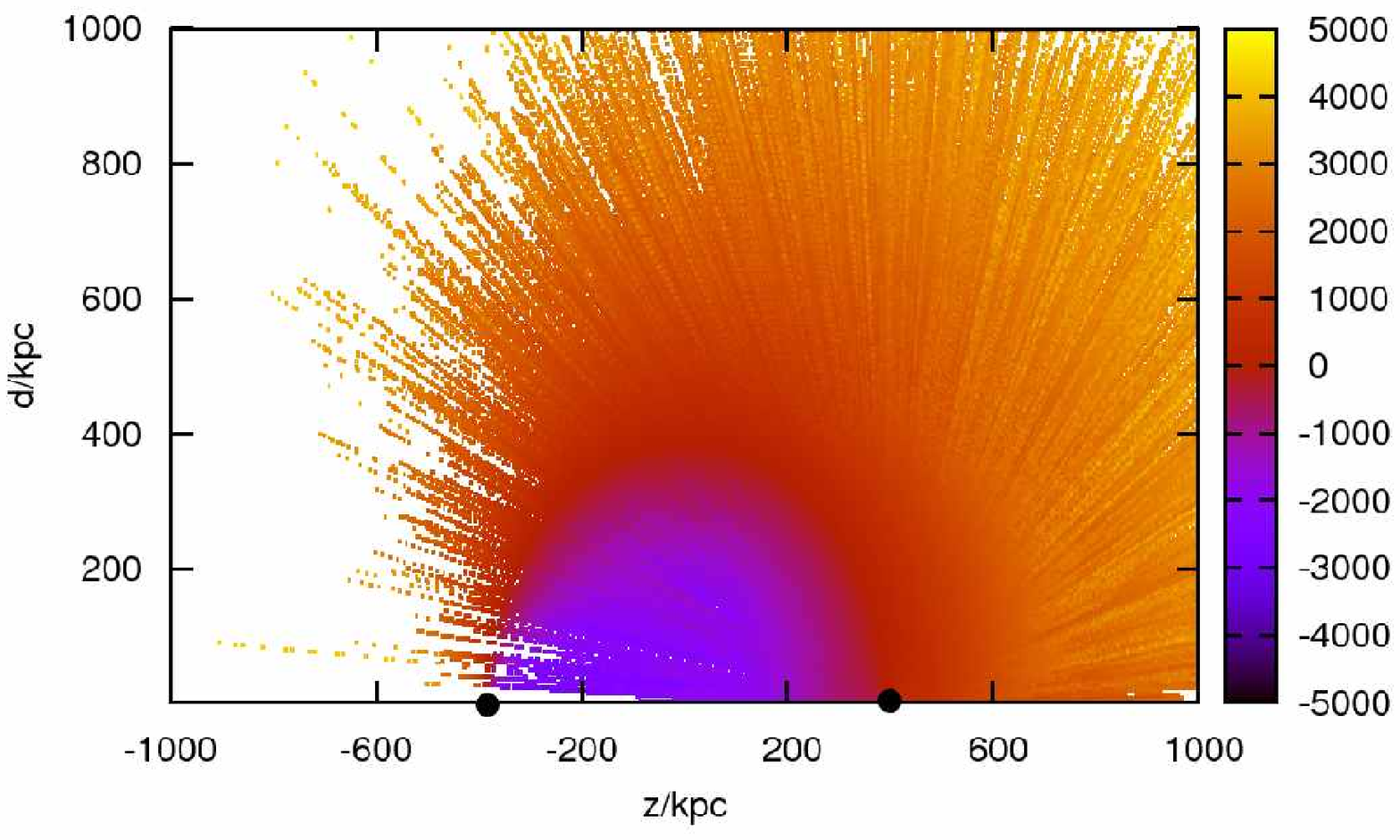}}
     \subfigure[average radial velocities in $\mathrm{km} \
     \mathrm{s}^{-1}$ (IBH)
     ]{\includegraphics[width=.48\textwidth,clip]{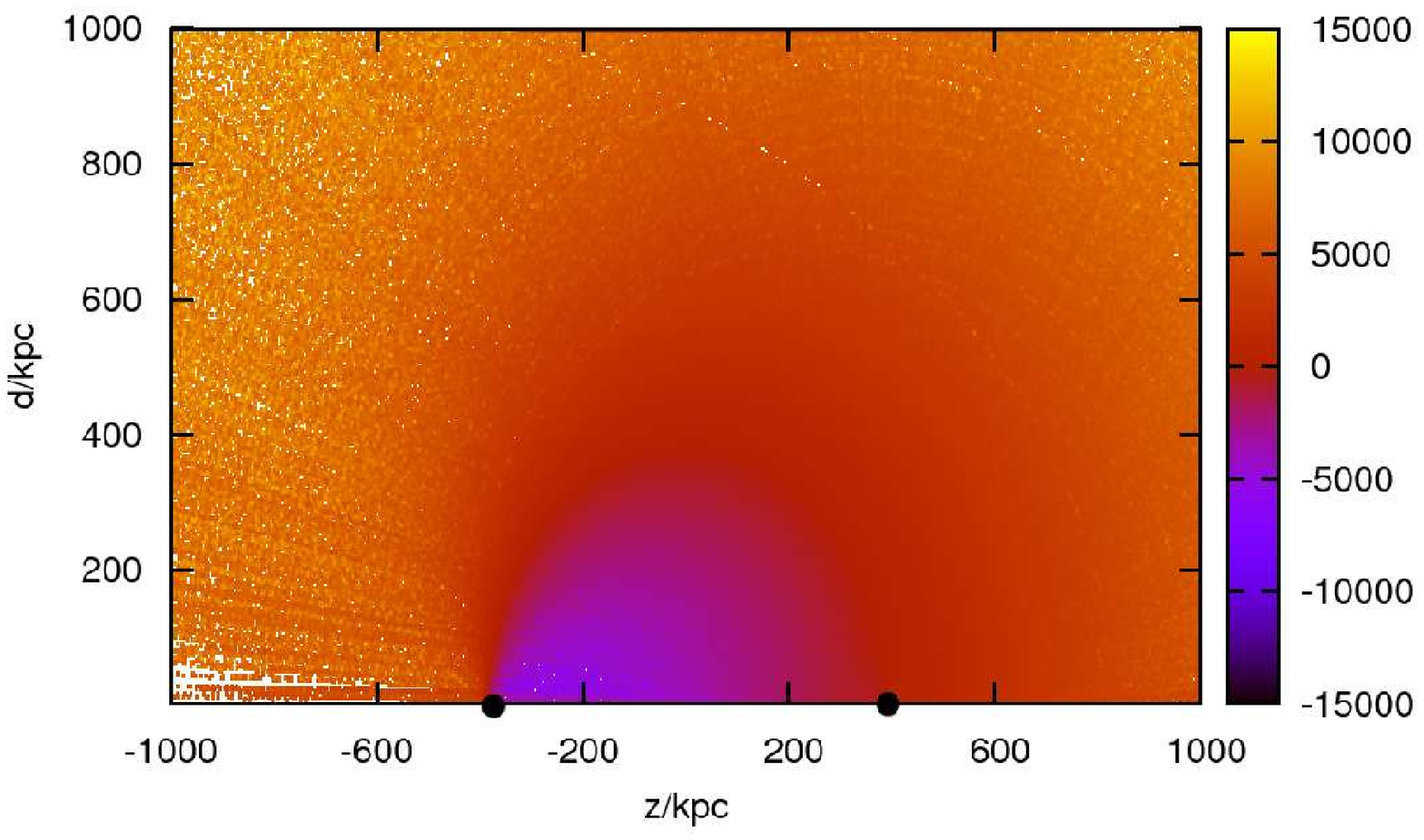}}\\
     \subfigure[average transverse velocities in $\mathrm{km} \
     \mathrm{s}^{-1}$ (TB)
     ]{\includegraphics[width=.48\textwidth,clip]{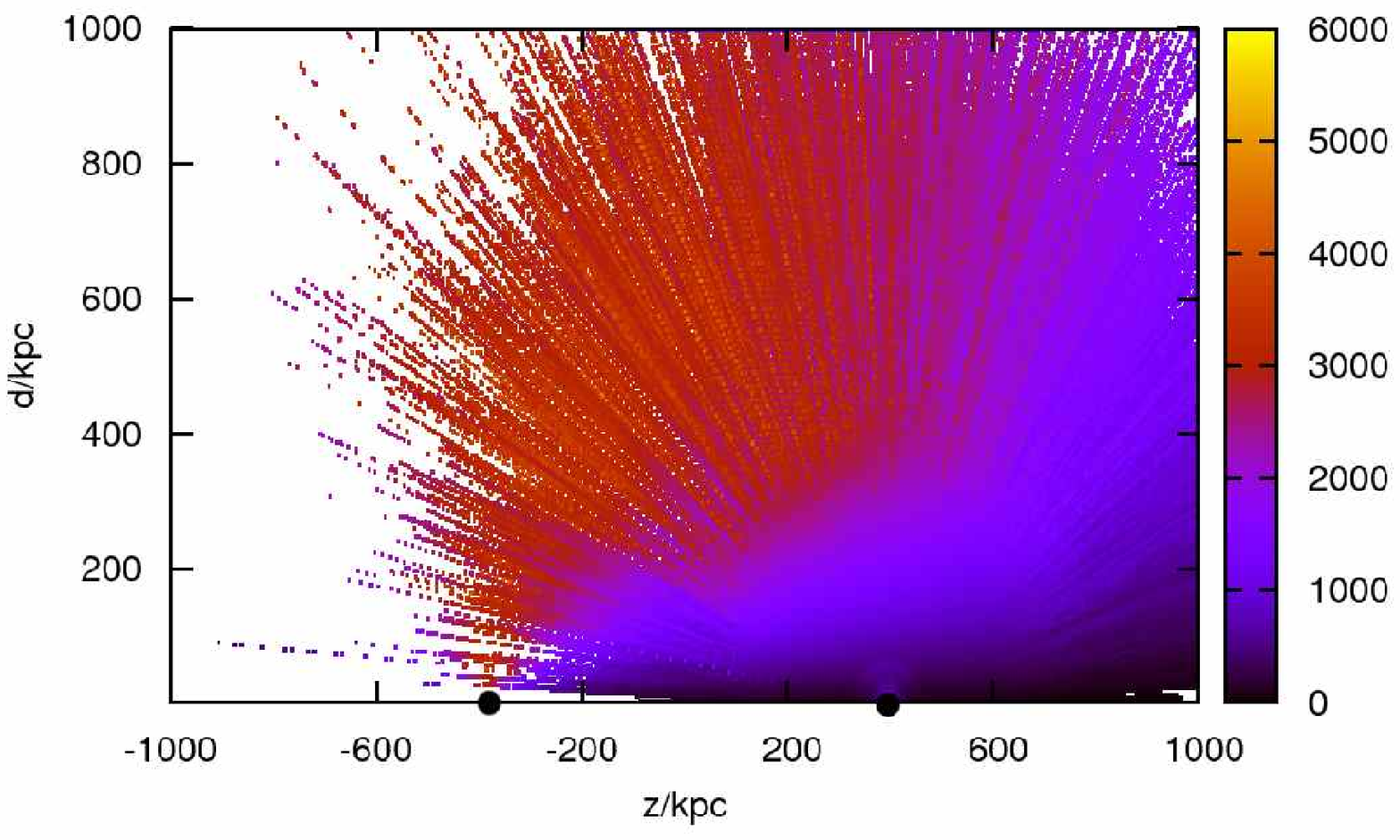}}
     \subfigure[average transverse velocities in $\mathrm{km} \
     \mathrm{s}^{-1}$ (IBH)
     ]{\includegraphics[width=.48\textwidth,clip]{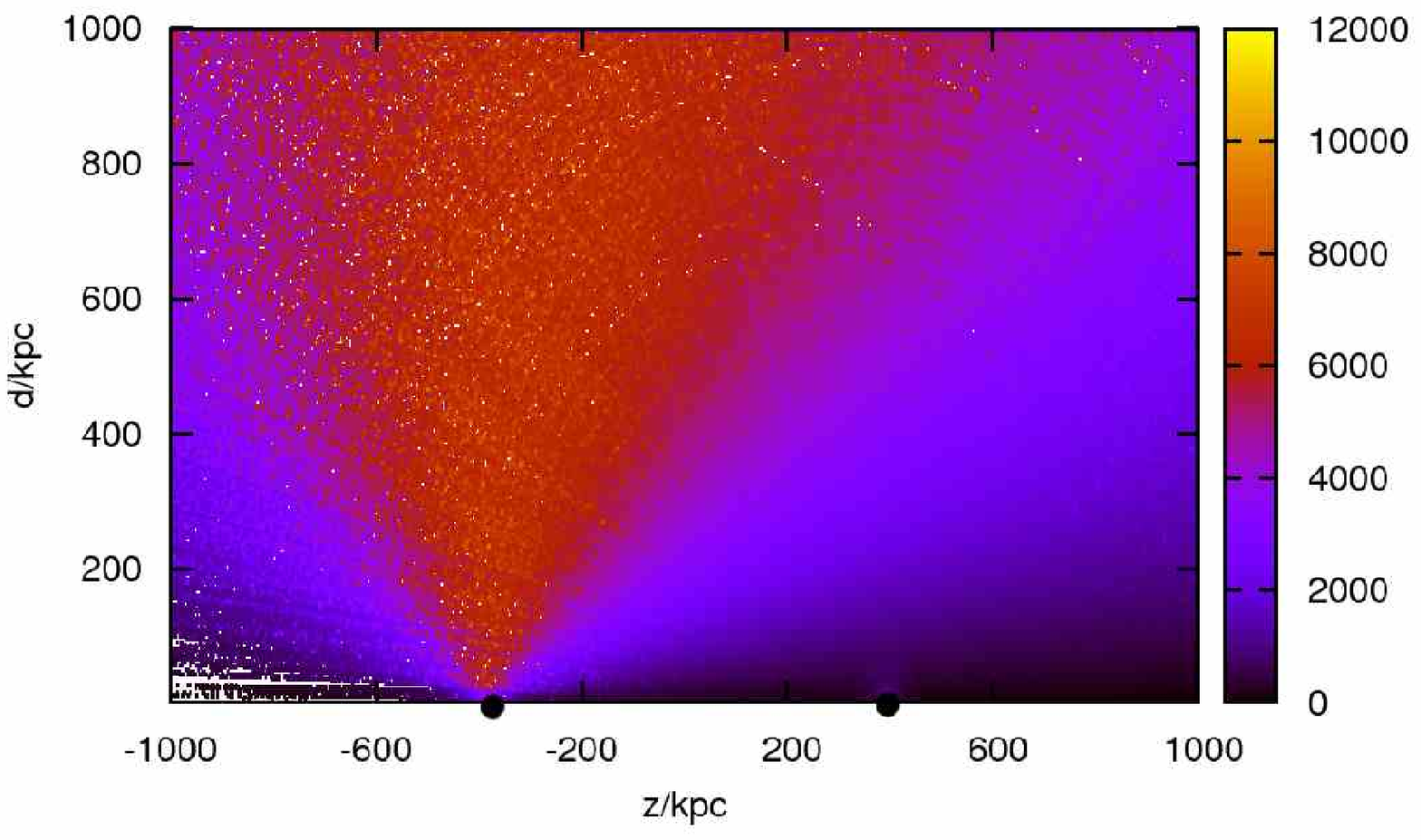}} \label{bluef}
     \caption{Same as Fig. 2 for stellar masses above $3 \msun$. }
\end{figure*}

The plots of the HVS distributions in Figures 2 and 3 exhibit a number of
interesting features. We first consider the TB mechanism mass bins (left
columns of Figs.\ 2, 3). It is clear that especially for the lower mass
stars, the HVS number density departs significantly from an isotropic distribution. For the low mass stars of the first bin, there is a sharp-featured
arc of increased number density ($n\sim 10^{-5}/$kpc$^3$, roughly
$10$ times the background density) which extends from M31 to the region
behind the MW. In addition, the star density is very high on axis
($\approx10^{-3}/$kpc$^3$). For higher mass stars, the deviations from
isotropy are smaller: for the third mass bin, the arc of raised
probability is hardly visible and the axial maximum is less pronounced; for the fifth and sixth mass bins, the HVS
distribution is almost entirely isotropic. Another feature which is 
(barely) visible in these plots is the presence of shell-like peaks in density
around M31. While shell-like features are often seen in galaxy merger
simulations (e.g.\ Hernquist \& Quinn 1989), changing our velocity
distribution showed that these shells were an artificial feature due to the
sharp cutoff in the lower ejection velocity. The density of the $\geq 3 \msun$ stars (Fig.\ 4) was
found to fall off quickly and isotropically with distance from M31. The
density is $10^{-6}/$kpc$^3$ at a distance of $\approx200$kpc from M31, and only
$10^{-9}/$kpc$^3$ at a distance of $\approx 500$kpc.

The lower-mass HVS distributions for the IBH mechanism (right columns of
Figs.\ 2, 3) are similar to the corresponding plots for the TB
mechanism, with only small differences. It can be seen that the arc
of first-mass-bin stars is less pronounced for the IBH mechanism, and that the IBH mechanism produces a somewhat lower HVS density in the MW halo. The distribution of the $\geq 3\msun$ IBH stars is
significantly different from the TB mechanism distribution. For the IBH mechanism, the
density of these stars falls off much more slowly with distance to give a value of
$10^{-9}/$kpc$^3$ at a distance from M31 of $\approx 800$kpc.

\subsection{HVS Velocities}
We now consider the stellar velocities measured relative to the MW centre. It must be noted when considering M31 HVSs close to the Earth that we have neglected the MW bulge and disc, so that the actual M31 HVS velocities there will be somewhat higher than we depict in our plots.

We first describe our results for the TB mechanism. Figure 2 shows that for the
first mass bin the radial and transverse speeds near the MW are typically $\approx
400$$\ \mathrm{km} \ \mathrm{s}^{-1}$. It is noticeable that the typical
speeds of higher mass stars near the MW were substantially greater. For the third
mass bin (Fig.\ 3) the typical velocity is $\approx500$$\ \mathrm{km} \
\mathrm{s}^{-1}$ near the MW; for the fifth mass bin it is
$\approx700$$\ \mathrm{km} \ \mathrm{s}^{-1}$. The high mass stars in Figure 4 appear to have
particularly high velocities of $\approx3000$$\ \mathrm{km} \
\mathrm{s}^{-1}$ near the MW, but of course the density of such stars is negligible in the regions where the speeds are so high. 

The velocities resulting from the IBH mechanism (Figs.\ 2, 3; right
column) were significantly higher than for the TB mechanism, with typical speeds near the MW of $600\ \mathrm{km} \ \mathrm{s}^{-1}$, $800\ \mathrm{km} \ \mathrm{s}^{-1}$, $1000\ \mathrm{km} \ \mathrm{s}^{-1}$ for the first, third and fifth mass bins respectively.

In all velocity plots the average velocities were larger at greater distances from the centre of M31 (though the numbers of HVSs there were smaller). The highest radial speeds occur
near the intergalactic axis, whereas the highest transverse speeds can be found perpendicular to this axis as expected.

\section{Discussion}

In our investigation, we found a chaotic sensitivity of the HVS trajectories to initial conditions. Chaotic regions of phase space are common for the motion of a test particle in a complicated three centre potential; chaos can even be found in non-stationary simple two body
potentials, such as binary systems (see Holman \& Wiegert 1999, Musielak et al.\ 2005).

The arc-like regions of raised density, which are seen for low mass stars
with both ejection mechanisms, are due to the focusing of ejected HVSs by
the MW's gravitational field. The stars that are focused in this way by the
MW have only just enough kinetic energy to escape the attraction of
M31. They are thus very slow and spend a large amount of time in the
observed ``arc'', which leads to an increased density there, before falling back towards the intergalactic axis. While one
expects regions of increased density, it is unclear why there is a
sharp-featured substructure within the arc, as there are no sharp steps in
the potential near this region. Such arc-like features are not observed for
higher-mass stars. This follows from the reduced main-sequence lifetime of
these stars ($t_l\propto m^{-2.5}$); high mass stars simply do not survive
long enough to be visible along such long-lasting trajectories. The arcs are less pronounced for the IBH mechanism because it
ejects HVSs at higher speeds, as can be seen from Equation (\ref{vi}).

It must be noted that the approximation of a static Local Group is not applicable to the longest-lasting trajectories. Taking the small initial separation of M31 and the MW into account would increase the number of low mass HVS from M31 we expect to find in the MW halo; the actual number densities of M31 HVSs in the MW may thus be higher than we calculate in Figure 2.

A large overdensity of stars was found near the intergalactic $z$-axis,
which is also due to gravitational focusing. As described before, the
gravity of the MW and the diffuse Local Group mass draw HVSs towards the
intergalactic axis; the axial symmetry of the entire configuration means
that every such trajectory eventually crosses the axis at some point, which
increases the density there. It is important to note that this
gravitational focusing has a very large effect; in Figure 2 it can be
seen that it increases the number density of stars near the the MW by
roughly \emph{three orders of magnitude} compared with the number of HVSs
at an equal M31-distance off axis.

Of course, the real Local Group mass distribution will not be
exactly axisymmetric. For instance, the symmetry of our model could be
broken by the galactic discs, any triaxiality of the MW or M31 halo (see Gnedin et al.\ 2005, Yu \& Madau 2007), or by
other Local Group galaxies. However, as the HVSs are usually moving very rapidly, their trajectories depend mainly on the large scale mass distribution. While local asymmetries may blur the axial overdensity somewhat, it is thus unlikely that they would eliminate this feature entirely.

The typical velocities of higher mass main-sequence stars are larger by selection; as more massive stars have shorter lifetimes, they require faster
speeds to travel a certain distance within their lifetime. Similarly, HVSs have larger velocities at greater distances from M31 because only a star with a large velocity can reach a location far from M31 within its lifetime. The increased
isotropy of the distribution of more massive stars is also due to the finite stellar lifetime: more massive live stars are faster, faster stars are deflected less
by the gravity of the Local Group, and less deflection implies a more
isotropic distribution.

The forms of the radial and transverse velocity distributions are as
expected for stars travelling away from M31. The shell-like minimum in the
transverse velocity $300$ kpc from M31 is due to many stars with velocities
just below the escape velocity turning around at this radius.

\section{Observational Outlook}
We may now examine the observability of HVSs from M31 based on the results of
our simulations. We first consider the large number of low mass ($\approx
1\msun$) stars expected within the MW halo, especially near the
intergalactic axis. The number density of HVSs in a large part of the MW halo is roughly $\sim 10^{-(3-4)}$/kpc$^3$ for the TB mechanism, and a few times less for the IBH mechanism. One therefore expects of order $\sim 1000$ HVSs from M31 in the MW halo for the TB mechanism, and a few hundred HVSs for the IBH mechanism. These stars have a number of features which
distinguish them from the background MW stars. First, the metallicity of
stars from galactic nuclei is different from stars in other regions of the
galaxy. Second, the HVSs from M31 have much larger typical speeds than
other stars in the MW halo. Their average radial speeds near the MW are at least
$400$$\ \mathrm{km} \ \mathrm{s}^{-1}$ (TB) and $600$$\ \mathrm{km} \
\mathrm{s}^{-1}$ (IBH); however, some of these stars will of course have
significantly larger speeds. In contrast to HVSs originating in the MW,
many M31 HVSs will be moving \emph{towards} the galaxy centre (on the side of
the MW which faces M31, which is where the solar system is located) with radial speeds
significantly exceeding the MW escape velocity. However, there is of course a contrast problem in
observing such M31 HVSs, as the MW contains a very large number of $\sim 1
\msun$ stars. A targeted spectroscopic search would probably be
unfeasible. However, it may be possible to detect these HVSs in planned
large scale surveys using telescopes such as Pan-STARRS, the \emph{Large Synoptic
Survey Telescope} (LSST) or the \emph{Global Astrometric Interferometer for
Astrophysics} (GAIA) satellite. GAIA will measure proper and
radial motions of $1 \msun$ stars out to at least $10$kpc (Perryman 2002, Yu \& Tremaine
2003) and should thus find a small number (of order $\sim10$) of
M31 HVSs. LSST will survey roughly half the sky and detect $1 \msun$ stars out to about $100$kpc (Claver et al.\ 2004, Sesana et al.\ 2007a); it should thus detect a few hundred M31 HVSs for the TB mechanism and approximately $\sim 100$ for the IBH mechanism. Even a few stars with sufficiently large negative radial velocities or transverse velocities would suggest a HVS production in M31. Such observations could be used to constrain the characteristics of the M31 galactic
centre environment as well as the Local Group mass distribution. One could
perhaps also distinguish between stars from the IBH and TB mechanisms by
their measured velocities; high velocity HVSs produced by the IBH mechanism would
indicate that the M31 MBH has an intermediate mass black hole companion
(which may be difficult to detect by other means).
\begin{figure}
\centering
\includegraphics[width=.49\textwidth,clip]{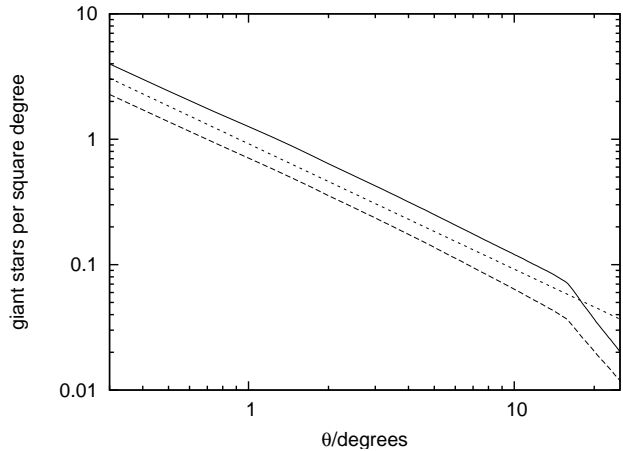}
\caption{Number of luminous hypervelocity RGB stars per square degree
within the distance range of $600$--$1000$kpc from the MW, as a function of
the angle from the centre of M31. Solid line: TB mechanism; long-dashed line: IBH mechanism; short-dashed line: for comparison, $0.92\times (\theta$/degrees$)^{-1}$ stars per square degree. One degree spans a distance of $13.6$ kpc near M31.}
\label{projblue}
\end{figure}

\begin{figure}
\centering
\includegraphics[width=.49\textwidth,clip]{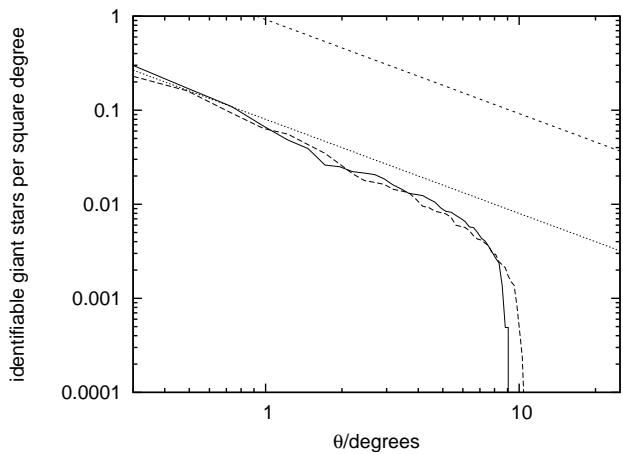}
\caption{Same as in Fig.\ \ref{projblue} but for identifiable hypervelocity RGB stars with radial approach velocities $<-540\ \mathrm{km} \ \mathrm{s}^{-1}$ (see text for details on velocity cutoff). Solid line: TB mechanism; long-dashed line: IBH mechanism; dotted line: for comparison, $0.08\times (\theta$/degrees$)^{-1}$ stars per square degree; short-dashed line: $0.92\times (\theta$/degrees$)^{-1}$ as in Fig.\ \ref{projblue}.}
\label{projbluevis}
\end{figure}

The current population of MW HVSs was discovered by observations of luminous main-sequence blue stars in the MW halo. From Figure 4, it is clear that the density of
$>3 \msun$ HVSs from M31 is only significant close to M31 due to the stars' short lifetime. Even massive $3-8 \msun$ main sequence HVSs would be too faint to be observed near M31 (unevolved $3 \msun$ blue have an apparent V magnitude of $\approx24$ at the distance of M31). Main sequence HVSs of masses greater than $\approx 8 \msun$ may be detectable in M31, but one would expect only $\sim 1$ such ejected star to exist at the present time (it would be close
to the M31 galactic centre). An observation of such a HVS would thus be very difficult.

However, there may be another possibility to detect more distant HVSs\footnote{We are grateful to Nelson Caldwell for this suggestion.}. Stars with masses over $1\msun$ brighten to $\sim10^{3} L_\odot$ for about $5\times10^6$ years when they reach the tip of the red-giant-branch (RGB) (Schaller et al.\ 1992); the luminosity of these stars would make them observable to very large distances. For instance, such a star near M31 has an apparent V magnitude of $\sim 21.5$ and is thus bright enough to investigate spectroscopically. We calculated the distribution of such giant HVSs (as for the main sequence HVSs) and found that they were only present at significant densities within $\sim 200$ kpc of M31 (as such stars are very rare due to the short duration of the bright phase, their densities are negligible far from their source). 

To examine the possibility of observing such RGB HVSs in the M31 halo, we constructed Figure \ref{projblue}, which depicts the density of hypervelocity red-giant-branch-tip stars per square degree as a function of the angular separation $\theta$ from M31 (along with an analytic fit to the results). The angular density of these stars is seen to fall off as $1/\theta$ at small angles; for $\theta>15^\circ$ the density drops off more rapidly due to the finite stellar lifetimes. Significant numbers of RGB HVSs can be found at angles beyond the galactic disc and bulge ($>1^\circ$), with of order $\sim 70$ (TB mechanism) or $\sim 30$ (IBH mechanism) hypervelocity RGB stars in the M31 halo at angles of $1^\circ \leq \theta \leq 10^\circ$. However, to distinguish such HVSs from background M31 halo stars and from MW foreground stars, they need to have radial approach velocities of $<-420 \mathrm{km} \ \mathrm{s^{-1}}$ (the velocity dispersion of halo stars is less than $\approx 140\ \mathrm{km} \ \mathrm{s^{-1}}$, see Chapman et al.\ 2006); it should be noted that this cutoff assumes a static Local Group and does not include the M31 approach velocity, so that the cutoff actually corresponds to a measured galactocentric radial speed of roughly $<-540 \mathrm{km} \ \mathrm{s^{-1}}$. The angular density of such identifiable RGB HVSs is depicted in Figure \ref{projbluevis}; their density is roughly a factor of ten lower, so that one could find of order $\sim5$ such giant stars within the M31 halo ($1^\circ \leq \theta \leq 10^\circ$). The background density of non-HVS RGB stars in M31 (Ibata et al.\ 2007, Fig.\ 18) is approximately four orders of magnitude greater than that of the HVS giants for $\theta>1^\circ$. A number of studies of the M31 halo population have recently been performed, in which spectra of thousands of M31 halo giants have been obtained (e.g.\ Chapman et al.\ 2006, Ibata et al.\ 2007). Hypervelocity red giant stars could be detected in similar future surveys of stars near M31.

There may be other sites aside from the MW and M31 where Local Group HVSs are produced. The star discovered by Edelmann et al.\ (2005) was recently studied by Bonanos \& Lopez-Morales (2007); the metallicity of this star is consistent with its suspected origin in the Large Magellanic Cloud. Due to this result, it seems likely that production of HVSs also takes place in the larger Local Group dwarf galaxies. In fact, an HVS production rate calculation for M32 by Lu et al.\ (2007) gave a rate comparable to the MW rate and higher than the M31 rate (though this calculation was only for the IBH mechanism). It is thus possible that in addition to the many low-mass M31 HVSs we predict to be located within the MW halo, there is also a significant population of HVSs from dwarf galaxies. This would increase the amount of information available from HVSs, though ambiguities in the origin of HVSs may make the analysis of the extragalactic HVS population more difficult. In some cases the origin of an HVSs could be determined from the speed and direction of its motion, but it may be difficult to distinguish, for instance, M31 and M32 HVSs. Of course, in order to make reliable predictions about HVSs from dwarf galaxies and determine whether they reach the MW halo in significant numbers, detailed calculations of dwarf galaxy HVS production rates and distributions must be performed. We hope to investigate dwarf galaxy HVSs in future work.

The use of extragalactic HVSs as probes of the Local Group mass distribution is certainly one of their most interesting applications. An investigation of what constraints these HVSs could impose on the mass distribution would be beyond the scope of this paper, but we hope to study this problem in the future.

\section{Conclusions}

In this paper we have investigated the distributions and velocities of the
hypervelocity stars from the M31 Galaxy by integrating a large number of
star trajectories within the Local Group. Below we summarise the main
conclusions from our calculations. It must be noted that, due to the uncertainties in the numbers of HVSs ejected and our simple Local Group model, the numerical values we give are only order-of-magnitude estimates.
\begin{itemize}

\item The differences between the galactic centres of M31 and the MW imply
that the BHC model for HVS production would not be effective in M31. The TB
mechanism and the IBH mechanism could produce large numbers of HVSs in M31.

\item Most HVSs ejected from M31 escape the local group, though some stars
with lower velocities are gravitationally drawn towards the MW. The HVS
trajectories show chaotic behaviour in some regions of phase space.

\item While high mass stars can only be found close to M31, there are large numbers of low mass ($\approx 1\msun$) M31 HVSs on the intergalactic axis and in an arc near the MW. These significant overdensities close to the MW are caused by gravitational focusing. We expect there to be roughly a thousand M31 HVSs within the virialized halo of the MW galaxy for the TB mechanism and a few hundred for the IBH mechanism. These HVSs from M31 could be distinguished from other stars by their large radial approach velocities ($\sim -500$$\ \mathrm{km} \ \mathrm{s}^{-1}$) and could thus be detected by future large scale astrometric surveys.

\item Our calculations of the number density of luminous hypervelocity RGB stars indicate that there are $\sim 30-70$ such stars in the M31 halo, of which $\sim 5$ should be identifiable and observable.

\end{itemize}

\section*{Acknowledgments} We are very grateful to Warren Brown for many helpful discussions and for his comments on the manuscript. We thank Nelson Caldwell for interesting discussions and thank Donald Lynden-Bell and Alberto Sesana for comments on the paper. BDS would like to acknowledge the support of the SAO intern program. This work was supported in part by NSF, Harvard University and FQXi funds.


\end{document}